\documentclass[11pt,letterpaper]{article}

\usepackage[utf8]{inputenc}
\usepackage[T1]{fontenc}
\usepackage{lmodern}
\usepackage[margin=1in]{geometry}
\usepackage{amsmath,amssymb}
\usepackage{booktabs}
\usepackage{array}
\usepackage{graphicx}
\usepackage{xcolor}
\usepackage{siunitx}
\usepackage[font=small,labelfont=bf]{caption}

\usepackage{tikz}
\usetikzlibrary{arrows.meta,positioning,calc,fit,backgrounds,decorations.pathreplacing,
                decorations.pathmorphing,shapes.geometric,patterns}
\graphicspath{{./}}

\definecolor{hlcell}{HTML}{C81E1E}
\definecolor{hlmdac}{HTML}{1B6FB5}
\definecolor{hlbias}{HTML}{0F8A54}
\definecolor{hldig} {HTML}{7A3FA0}
\definecolor{hlaux} {HTML}{4A4A4A}
\definecolor{hlresc}{HTML}{D07A00}

\usepackage{pgfplots}
\pgfplotsset{compat=1.18}

\usepackage[colorlinks=true,linkcolor=black,citecolor=black,urlcolor=blue]{hyperref}

\usepackage[backend=biber,style=nature]{biblatex}
\hypersetup{
  pdftitle={Composability rather than computation sets the cost of an analog
            EML hardware fabric},
  pdfauthor={Christof Teuscher},
  pdfsubject={Unconventional computing / analog hardware}}

\makeatletter
\let\eml@textsc\textsc
\DeclareRobustCommand{\textsc}[1]{%
  \edef\eml@ser{\f@series}%
  \ifx\eml@ser\mddefault \eml@textsc{#1}\else \MakeUppercase{#1}\fi}
\makeatother

\providecommand{\gap}[1]{}
\newcommand{\note}[2]{\section{#1}\label{#2}}

\title{\bfseries Composability rather than computation sets the cost
  of an analog \textsc{eml} hardware fabric}

\usepackage{orcidlink}

\author{Christof Teuscher\,\orcidlink{0000-0002-5927-1900}\\
\small Department of Electrical and Computer Engineering\\
\small Portland State University, Portland, OR 97201, USA\\
\small \texttt{teuscher@pdx.edu}\\
\small ORCID: \href{https://orcid.org/0000-0002-5927-1900}{0000-0002-5927-1900}}

\date{\today}

\begin{document}
\maketitle

\begin{abstract}
\noindent
The operator $\mathrm{eml}(x,y)=\exp(x)-\ln(y)$ with the constant $1$
generates the elementary functions, a continuous counterpart to
\textsc{nand}. Whether it yields a useful fabric had not been asked of
hardware. We ask in network models, circuit simulation and SkyWater
\SI{130}{\nano\meter} layout. Four bipolar junctions evaluate the
operator for \SI{13}{\femto\joule}, beating a width-matched digital
datapath by \numrange{4}{134}$\times$. The fabric assembled from them
is not cheap: it loses to resource-matched baselines, and over the
reals its grammar excludes trigonometry. Amplifiers holding those
junctions' operating points take \SI{74.5}{\percent} of a cell's
current, so a cell costs \num{3000} times what they spend. Extracted
non-idealities cost \num{2.6}$\times$ when a cell must hold
a value and nothing when it need only be repeatable. Sharing them
across cells recovers two of the three orders. The premise was
that a universal primitive licenses a uniform machine. It survives in the
primitive and fails in the machine.
\end{abstract}

\noindent\textbf{Keywords:} analog computing, unconventional computing,
functional completeness, elementary functions, translinear circuits,
energy efficiency, SKY130

\clearpage

\section{Introduction}
\label{sec:intro}

The elementary functions form a heavily redundant basis. Arithmetic,
the exponential and the logarithm, powers and roots, and the circular
and hyperbolic functions with their inverses can be interderived in
many ways, and the classical reductions shrink that basis without
collapsing it to a single generator. The exponential--logarithmic
representation and Euler's formula are the familiar
ones. Odrzywo\l{}ek has shown, by systematic search over candidate
binary operators, that one continuous operator does collapse
it~\cite{odrzywolek2026eml}:
\begin{equation}
  \mathrm{eml}(x, y) = \exp(x) - \ln(y).
  \label{eq:eml}
\end{equation}
Together with the constant $1$, \eqref{eq:eml} generates arithmetic, the
standard transcendental and algebraic functions, and the constants $e$,
$\pi$ and $i$. Every expression the construction reaches is a binary tree
of identical nodes produced by the grammar
$S \rightarrow 1 \mid \mathrm{eml}(S,S)$. In the parameterized form used
by Odrzywo\l{}ek, each port input of each node is an affine combination
$\alpha + \beta x + \gamma f$ of an external variable and an upstream
node value, so a tree of fixed shape becomes a family of expressions
indexed by continuous weights.

A single primitive from which an entire algebra is generated by
composition is, structurally, what the Sheffer stroke is to Boolean
logic~\cite{sheffer1913}. The mathematical content of Sheffer's result
is a completeness theorem; its consequence for hardware was larger and
different in kind. If one cell suffices, a designer characterizes that
cell once. Its device sizes, its timing model, its test pattern and
its yield statistics are established once, and the cell is then
replicated. A universal primitive appears to license a kind of machine
rather than a smaller component count: uniform, tiled, characterized
once and instantiated everywhere.  That expectation is the hypothesis
we test here, and the precise claim available to support it is
narrower than the analogy suggests. Equation \eqref{eq:eml} is
functionally complete for a class of elementary functions in the same
sense that \textsc{nand} is functionally complete for the Boolean
functions. Functional completeness for a function algebra is not
Turing-universality. An $\mathrm{eml}$ tree is a finite, feed-forward,
clockless expression evaluator with no state, no branching and no
unbounded memory, and no statement in this paper should be read as a
claim of computational universality in the Turing sense.

We test that expectation mostly in simulation. We characterized the cell
and the chain by circuit simulation on the netlist matching the submitted
layout and on a parasitic-annotated extraction of it, enumerated the
grammar over an expression corpus, and compared a fabric of such cells
against resource-matched software and digital baselines. Underneath that
study sits a translinear $\mathrm{eml}$ cell we designed in the open
SkyWater \SI{130}{\nano\meter} process~\cite{skywater2022pdk}, in which
junction physics delivers the exponential and the logarithm and
Kirchhoff's current law performs the subtraction in \eqref{eq:eml} at a
summing node; we assembled two of them into a chain with programmable
weighted interconnect, completed physical design and verification, and
submitted the result to a Tiny Tapeout shuttle~\cite{venn2024tinytapeout}. The
build is not the subject here. It supplies the device-derived parameters the
network study runs on, and it shows that the operator is physically
realizable.

The primitive is as cheap in silicon as the completeness result
suggests. Four bipolar junctions evaluate \eqref{eq:eml}; they occupy
\SI{2.7}{\percent} of the shipped cell and, run bare, would deliver an
evaluation for \SI{13}{\femto\joule}. The fabric built around them is
not usable. It fails on every axis we tested. Its grammar is closed
to trigonometry, which excludes \num{18} of the \num{100} AI Feynman
equations outright, while only \num{11} of the remaining \num{82}
contain $\exp$, $\log$ or $\tanh$ at all and only \num{10} contain
$\exp$ or $\log$. Against baselines matched to its own parameter count
it (1) loses to a multilayer perceptron by
\numrange{7.4}{20.9}$\times$ (median \num{11.7}$\times$ over the five
synthetic targets any model fitted), (2) wins none of sixteen
closed-loop dynamical-system tasks, and (3) is beaten on sensor
linearization by a four-parameter polynomial that a
\num{92}-parameter fabric with ideal hardware does not match. Its energy
per evaluation exceeds a digital datapath of the same precision by one
to two orders of magnitude at every point of a sensitivity box spanning
the plausible digital-modeling assumptions.

The mechanism behind those failures is a single quantity. It is the
result we consider transferable. The energy the fabric spends per cell
exceeds the energy its own translinear core spends by
\num{3046}$\times$. It is a per-cell quantity and it does not amortize:
a seven-cell fabric pays it seven times rather than diluting it. That
ratio is $E_\mathrm{fab}/E_\mathrm{core}$, and it contains no digital
model, no node-scaling factor and no control overhead, so it is
invariant across the sensitivity box by construction rather than by
good fortune. The fabric spends its charge on
holding the primitive's operating points, making its analog value
addressable, and storing its per-cell calibration. None of it goes to
computing. Sarpeshkar places the analog advantage at low
precision~\cite{sarpeshkar1998analog}; we reproduce his crossover at
\num{13.8} bits and find the fabric operating at \numrange{2.28}{5.73}
bits, nine to eleven bits inside the regime where analog should win, and
losing anyway. The reconciliation is that both terms in the fabric's
energy law are ``scaffolding'' terms and $\mathrm{eml}$ contributes
neither. A universal primitive can be nearly free in silicon and still
yield an unusable fabric, because composability rather than computation
sets the cost.

We submitted the design to a Tiny Tapeout SKY26c shuttle, but no part
has been returned and no number in this paper is a measurement of a
physical device. Throughout, \emph{simulated} denotes a quantity we
obtain by circuit simulation (\texttt{ngspice}~\cite{ngspice} with the
sky130A model library) from the netlist that matches the submitted
layout under layout-versus-schematic comparison, or read as a bounding
box off that layout; \emph{post-layout} denotes a quantity taken on a
parasitic-annotated extraction of the same database, which we
performed for the single cell and for the two-cell assembly and for no
deeper structure; and \emph{modeled} denotes a construction whose
parameters are simulated in the preceding sense but whose functional
form is assumed. This is a simulation and physical-design study, and
every quantitative claim below states which of these three categories
it belongs to.

The remainder of this introduction presents the operator and its grammar, the
port asymmetry that shaped the cell design, and the work's relation to
translinear and log-domain analog circuits, reconfigurable analog arrays,
digital elementary-function hardware, symbolic regression, and the
analog-versus-digital energy argument the final result must be reconciled with.
Results then reports the cell characterization, the expressivity boundary, the
baseline comparisons, the injection of hardware-derived non-idealities into the
behavioral model (Section~\ref{sec:res-nonideal}), where the silicon stops
being the subject and becomes a source of parameters, and the energy law and
the composability overhead, the energy a complete cell spends per evaluation
divided by the energy spent by the four junctions inside it that evaluate the
operator. The Discussion states what the negative result establishes, what it
does not, what would have to change for the approach to work, what substrate
property that change would require, and the limits of a study conducted
entirely in simulation and physical design. Methods, which follows the
Discussion, describes the cell, the two-cell assembly, the physical-design and
verification flow, the simulation and network-modeling procedures, and the
construction of the resource-matched baselines.

Odrzywo\l{}ek's generative claim is that the expressions produced by
$S \rightarrow 1 \mid \mathrm{eml}(S,S)$, with each port input an affine
combination of a variable and an upstream node value, contain the elementary
functions~\cite{odrzywolek2026eml}. Watch the grammar compute once. The constant and the exponential each fall out of a single node,
$\mathrm{eml}(1,1) = e$ and $\mathrm{eml}(x,1) = e^{x}$. The logarithm costs
three, $\ln z = \mathrm{eml}(1,\mathrm{eml}(\mathrm{eml}(1,z),1))$, which
unwinds from the inside. What a tree costs in hardware is set by that depth,
and depth is what the rest of this paper is about.

Two features of the claim matter for hardware. The first is uniformity: every
internal node is the same two-argument operation, so a machine implementing
\eqref{eq:eml} with a weighted interconnect implements the whole reachable
set, and expression selection reduces to setting weights and wiring. The
second is domain. Over the reals the logarithm restricts its second argument
to positive values, and the restriction compounds with depth because that
argument is a subtree output. %
The circular functions appear in the
construction only through complex arguments: Odrzywo\l{}ek states that the
internal computations for the trigonometric functions in particular must be
carried out in the complex domain, even where the compiled expression is
evaluated on the real axis~\cite{odrzywolek2026eml}. That places them outside
a real-valued current-domain machine. That exclusion is a property of the
algebra and not of the circuit, and we quantify in Results the fraction of a
physics-equation corpus it removes.

The operator's two ports are not interchangeable, and the asymmetry between
them shaped both the cell design and the analysis. Write
$f = \mathrm{eml}(u,v) = \exp(u) - \ln(v)$. The partial derivatives
are
\begin{equation}
  \frac{\partial f}{\partial u} = \exp(u),
  \qquad
  \frac{\partial f}{\partial v} = -\frac{1}{v},
  \label{eq:ports}
\end{equation}
the first proportional to the quantity being computed and the second
inversely proportional to it. An error presented at the exponential port
is amplified by the magnitude of the value it corrupts; an error
presented at the logarithmic port is divided by it. The two arguments of
$\mathrm{eml}$ therefore behave, from a hardware standpoint, as two
different circuits sharing a cell.

Under composition the asymmetry becomes qualitative. A chain through the
$u$ port iterates $\exp$ and is doubly exponential in depth, so it
consumes any finite device span within a small number of stages and
converts device mismatch into a log-normally distributed output carrying
a mean bias. A chain through the $v$ port iterates $\ln$, which contracts
error towards a fixed point at any depth, but which discards signal
amplitude at every level, so depth in that direction is limited by signal
loss rather than by error growth. Equation \eqref{eq:ports} is a property
of the operator: $\exp(u)$ is unbounded above, and $-1/v$ carries a pole
at the origin. No better amplifier, finer trim or different process
removes it, and it is the reason a cell built around \eqref{eq:eml} must
be held to a precision that its own arithmetic never uses. The asymmetry
shaped how we designed the cell, and we expected it to be the property
that bounds what a fabric of such cells can do. Device mismatch got there
first. We take that up in the Discussion.

Gilbert's translinear principle states that around a closed loop of
forward-biased junctions the product of current densities in one direction
equals the product in the other, which yields products, quotients and powers
directly from device physics~\cite{gilbert1975translinear,gilbert1996historical}.
A forward-biased bipolar junction is exponential in its base--emitter voltage
by Boltzmann statistics over many decades of collector current, without
approximation, iteration or stored coefficients; the same device inside the
feedback loop of an amplifier that holds its collector fixed inverts the
relation and delivers the logarithm. Subthreshold CMOS obeys the same
law~\cite{vittoz1977weak,mead1989analog}, and log-domain and companding filters
exploit it to obtain linear signal processing from exponential
devices~\cite{frey1993logdomain,seevinck1991companding}. The surrounding
practice for biasing, mirroring and matching is
standard~\cite{razavi2017analog}, and matching follows the area law of Pelgrom
et al.~\cite{pelgrom1989matching}.

This substrate is unusually well matched to \eqref{eq:eml}. Both of the
operator's transcendental operations are delivered by junction physics
rather than approximated, and its only other operation is a difference,
which in the current domain is Kirchhoff's current law at a shared node.
The operator therefore maps onto a small number of bipolar devices and a
summing node. The same exponential law also fixes the cost of everything
built around it: a node servoed to within $\delta V$ of its target
produces a relative current error of $\exp(\delta V / n V_T) - 1$, so
accuracy at the milli-volt scale of the thermal voltage is
percent-scale accuracy in current. A translinear $\mathrm{eml}$ cell is
thus a voltage-reference problem with exponential sensitivity, and the
amplifiers that hold its operating points dominate its cost, not the
devices that compute. What is new here is not the
circuit style but the function the circuit computes, the intention to
tile it, and the accounting that intention forces.

Field-programmable analog arrays are the closest existing hardware in
structure: tiled configurable analog blocks joined by programmable
interconnect, demonstrated at large scale in standard CMOS and reviewed by
Hasler~\cite{hasler2007large,schlottmann2012fpaa,hasler2020fpaa}. The
difference is not that their blocks are restricted to linear operations.
They are not: a computational analog block typically carries transistors,
transconductance amplifiers, passives and more complicated elements such as
multipliers, and such devices have been used to solve nonlinear differential
equations. The difference is in where the freedom sits. An FPAA block is
assembled from a component library at configuration time, so the tile's
function is chosen by the designer; a fixed-primitive fabric inverts that,
taking one non-configurable operation chosen by a completeness result and
leaving only the interconnect weights free. An $\mathrm{eml}$ array is
therefore not a retuned FPAA.

One feature of the mature FPAA literature bears directly on the accounting
this paper sets up, and it cuts against the expectation we began with. In
floating-gate-enabled devices the programmable interconnect is not overhead
that computes nothing: the same floating-gate elements that act as routing
switches also store analog values and perform vector--matrix multiplication
in the routing fabric, so that, as Hasler puts it, switches are not dead
weight~\cite{hasler2020fpaa}. That is an architectural answer to the question this paper ends on. If the machinery that makes a fabric composable can itself compute, the distinction between computation and scaffolding stops being a subtraction. 

Before any claim is made about generality, the other tiled analog
architectures belong beside the \textsc{fpaa}. Table~\ref{tab:fabrics} puts
them there, on the single axis this paper is about: what each one spends on
computing, against what it spends on making that computation composable.

\begin{table}[tb]
\caption{\textbf{Tiled analog architectures, and what each spends on
composability.} The last column is sparse because most of this literature
reports total power rather than the split between the computing element and
the machinery that holds, addresses and calibrates it. That is the measurement this
paper asks the field to start reporting, and the reason the present work
gives a ratio rather than a power. The \textsc{fpaa} row is empty for a
different reason: in floating-gate devices the routing fabric computes, so
the division the column assumes is not the division that architecture
makes.}
\label{tab:fabrics}
\centering\small
\begin{tabular}{@{}>{\raggedright\arraybackslash}p{0.20\linewidth}p{0.20\linewidth}p{0.26\linewidth}p{0.24\linewidth}@{}}
\toprule
Architecture & What computes & What makes it composable & Reported split \\
\midrule
\textsc{fpaa}~\cite{hasler2007large,schlottmann2012fpaa,hasler2020fpaa}
  & configurable analog blocks: transistors, amplifiers, passives,
    multipliers
  & per-tile programming and routing, though in floating-gate devices the routing also computes
  & the split does not apply in the same form \\
\midrule[\cmidrulewidth]
Crossbar in-memory~\cite{shafiee2016isaac,murmann2021mixedsignal,haensch2019next}
  & the resistive array performing multiply--accumulate
  & converters and drivers, \emph{shared across many rows}
  & array \SI{2.4}{\milli\watt} against \SI{16}{\milli\watt} for the
    converters: \SI{10}{\percent} of unit power, \num{9}$\times$ \\
\midrule[\cmidrulewidth]
Continuous-time hybrid~\cite{guo2016hybrid}
  & analog integrators
  & conversion and digital control
  & same division reported \\
\midrule[\cmidrulewidth]
Analog \textsc{kan}~\cite{escudero2026akan}
  & reconfigurable nonlinear-processing units
  & transimpedance amplifier, \textsc{dac}s and \textsc{adc}s
  & units \SI{50}{\nano\watt}, amplifier \SI{94}{\micro\watt} \\
\midrule[\cmidrulewidth]
This work
  & four bipolar junctions evaluating $\mathrm{eml}$
  & servo amplifiers, bias, output mirrors, \emph{per cell}
  & \num{3046}$\times$ per cell \\
\bottomrule
\end{tabular}
\end{table}

The crossbar row is worth reading against ours in detail, because the two are
the closest quantitative comparison available and they disagree by two orders
of magnitude for a reason. In \textsc{isaac}'s modeled \SI{32}{\nano\meter}
tile the memristor array that performs the multiply--accumulate draws
\SI{2.4}{\milli\watt} and occupies \SI{1.5}{\percent} of the unit's area,
while the analog-to-digital converter alone draws \SI{16}{\milli\watt} and
occupies \SI{73}{\percent} of it; converters and sample-and-holds together
are \SI{83}{\percent} of unit power. The computing element is thus about a
tenth of the power and a sixtieth of the area, and the peripheral-to-compute
power ratio is roughly \num{9}.

Set beside this work, the agreement is closer than the disagreement. Our
translinear core occupies \SI{2.7}{\percent} of its cell against \textsc{isaac}'s \SI{1.5}{\percent}. In both cases the arithmetic is a rounding error in the floorplan. What differs is the ratio: \num{9}$\times$
there against \num{3046}$\times$ here, and the cause is in the third column.
A crossbar shares eight converters across eight arrays of \num{128} rows, so
the conversion cost is divided by the array before it reaches any one
multiply--accumulate, and it falls further as the array grows. A fabric whose
cells must each be held at their own operating points divides by nothing.
Every row of the table has the same shape, a cheap computing element beside expensive peripheral machinery. The word that separates them is \emph{shared} against \emph{per cell}. That distinction, rather than the
existence of the tax, is what this paper is about.

The established alternative is digital. CORDIC evaluates the
trigonometric, hyperbolic and related functions by shift-and-add
iteration~\cite{andraka1998cordic}, and Detrey and de Dinechin's
polynomial and table-driven operator generators produce pipelined
elementary-function cores at a specified
precision~\cite{detrey2007flopoco}. Their cost is incurred per
function, per iteration and per bit, and the designer buys accuracy with
area and latency; their decisive advantage is that accuracy composes
without calibration, since a correctly rounded result carries no
per-instance offset into the next stage. Horowitz has characterized
their energy well at the level of individual arithmetic and memory
operations~\cite{horowitz2014computing},
which makes a like-for-like comparison possible without synthesizing a
core, at the price of a modeling assumption that must be carried
explicitly. These are pipelines rather than tiles: the hardware is
specialized to a function, and evaluating a different function means
instantiating different hardware. The comparison an analog primitive
must survive is against this alternative at matched precision, and
against the simplest form of it a designer would build.

Machines that evaluate and compose elementary functions matter now
because several active lines of work search that space. Symbolic regression recovers closed-form expressions from data by
searching a discrete space of expression trees; Udrescu and Tegmark
search it by physics-guided decomposition~\cite{udrescu2020aifeynman},
Cranmer et al. with learned priors~\cite{cranmer2020discovering}. Liu et
al. replace fixed activations with learnable univariate functions placed
on network edges, so that the trained Kolmogorov--Arnold network is a
composition of one-dimensional functions and admits symbolic
readout~\cite{liu2025kan}. In each case the
computational load is the repeated evaluation and composition of
elementary functions under continuous parameters. Odrzywo\l{}ek's affine
port parameterization gives an $\mathrm{eml}$ tree the same structure:
fixed topology, continuous trainable weights, and a converged weight set
that reads out as a formula~\cite{odrzywolek2026eml}. A fabric of
$\mathrm{eml}$ cells with programmable weighted interconnect is therefore
a candidate substrate for this class of computation, and the AI Feynman
equation set~\cite{udrescu2020aifeynman} supplies the natural test of
whether the operator's grammar covers the functions such a search would
have to reach.

Sarpeshkar made the argument that analog computation is efficient at low
precision and loses at high precision~\cite{sarpeshkar1998analog}.
In a current-mode circuit the charge that must be delivered to resolve
$b$ bits against shot noise grows as $4^{b}$, while a digital datapath
grows polynomially in $b$, so the two curves cross and analog is the
cheaper machine below the crossover. The argument is about a computing
element in isolation. It says nothing about what it costs to make that
element addressable, to hold it at an operating point, to compose its
output with another element's, or to store the calibration that mismatch
between nominally identical elements forces. Those costs are the subject
of the present work, and the reconciliation between a reproduced
crossover and a fabric that loses well inside the analog regime is the
central result of the Discussion.

Analog computing has drawn sustained renewed interest, and researchers
have pursued it on two fronts: as a reconstruction of the classical
analog machine in modern
technology~\cite{ulmann2022analog,ulmann2024beyond}, and as part of a
broader program of computing with physical
dynamics~\cite{markovic2020physics,teuscher2014unconventional}. In much
of that work the designer begins with a substrate,
a reaction--diffusion medium or a memristive film or a disordered dopant
network, and asks what it can be made to compute, so the
resulting computational vocabulary is whatever the material offers. We
run the arrow in the opposite direction. The operator is fixed in
advance by a completeness result, and the design question is which
physical mechanism delivers that specific operator and at what cost. We are not aware of a
previously reported array whose native operation is a universal
transcendental primitive, and that object, rather than the individual cell, is
the subject of this paper.

\section{Results}
\label{sec:results}

\subsection{The grammar excludes trigonometry and rarely matches the corpus}
\label{sec:res-grammar}

We enumerated the expression grammar semantically, so that two trees are the
same entry if and only if they compute the same function on the evaluation
grid. This gave \num{44559} distinct functions at depth $\le 4$ from
\num{2090918} syntactic trees. Of the \num{194688} candidate depth-4
compositions, \num{148016}, or \SI{76.0}{\percent}, are out of domain: the
logarithm's argument goes non-positive somewhere on the grid, or the
exponential overflows. That fraction is a property of $\mathrm{eml}$ and
appears before any hardware error is applied.

Over the reals the closure of $\{1, x\}$ under $\exp$, $\ln$ and subtraction
is the Liouvillian exponential--logarithmic tower, which contains no
trigonometric function. Consistent with this, $\pi$, $\sin$, $\cos$, $\tan$,
$\arctan$, $\sinh$, $\cosh$ and $\tanh$ returned \num{0} hits across the
\num{44559} exhaustively enumerated classes and a further \num{2.96e6}
targeted ones. The boundary is visible in training as well as in enumeration.
On ideal hardware $\sin(3\ln x)$ is learnable (median RMSE \num{0.127} at
depth 4, best \num{0.027} at depth 6, $n = 3$ seeds, target RMS \num{0.690}),
while we learned neither $\sin(5\ln x)$ nor $\sin(8\ln x)$ at any depth,
initialization or budget attempted: RMSE stayed at \numrange{0.61}{0.68}
against a target RMS of \num{0.71}. Trigonometry is a hard boundary of the
grammar, not a difficulty within it.

We then applied the same test to a physics corpus. We assembled the
\num{100}-equation AI Feynman set and verified every entry by structural
consistency, SI dimensional homogeneity, sampler finiteness and hand-written
physics checks (\num{41} checks run, \num{0} failed, \num{54} equations
carrying at least one independent check; the remaining \num{46} rest on
structure and dimensions alone). Of the \num{100}, \num{18} are blocked by an
irreducible trigonometric factor and are therefore unrepresentable in this
grammar at any size. That exclusion is not specific to $\mathrm{eml}$: as the Discussion
sets out, no real primitive whose closure
lies in the exponential--logarithmic--algebraic--$\arctan$ class represents
trigonometry exactly, so the same \num{18} are blocked for every operator in
the class. Of the \num{82} that remain, only \num{11} contain
$\exp$, $\log$ or $\tanh$ at all, and only \num{10} on the narrower basis of $\exp$ and $\log$ alone. The operator census over the corpus is
\texttt{mul} \num{100}, \texttt{div} \num{77}, \texttt{add} \num{47},
\texttt{pow} \num{45}, \texttt{sqrt} \num{18}, \texttt{cos} \num{10},
\texttt{exp} \num{9}, \texttt{sin} \num{8}, \texttt{asin} \num{2},
\texttt{log} \num{1}, \texttt{tanh} \num{1}. The primitive's native operations
are needed by roughly one target in nine, and the operation it cannot perform
appears in nearly one in five. Of the \num{71} purely algebraic equations
\num{38}
are bare monomials, products and quotients of powers containing no addition at
all, for which the grammar offers nothing a multiplier does not.

Depth in the exponential direction is also self-limiting. Defining $u$-depth
as the maximum number of first-argument edges on any root-to-leaf path,
\SI{97.0}{\percent} of the \num{44559} classes sit at $u$-depth 3, but the
fraction that stays in the simulated device span and remains non-constant is
\SI{90.0}{\percent} at $u$-depth 1, \SI{95.8}{\percent} at $u$-depth 2 and
\SI{12.3}{\percent} at $u$-depth 3. The third exponential hop removes
\SI{87.7}{\percent} of the classes it creates. In the joint distribution over
total depth and $u$-depth, \SI{96.4}{\percent} of the corpus sits in the single
cell $(d,p) = (4,3)$, so cell count is not the binding cost.

\subsection{Resource-matched baselines win at every size tested}
\label{sec:res-matched}

We match every comparison in this subsection by parameter count, and we say
so at each one. We trained the fabrics with Adam. Where a validation split
exists, baseline hyperparameters are selected on it: the matched multilayer
perceptron is chosen over hidden-layer count, activation and learning rate on
a held-out quarter of the training split and then retrained on the whole of
it, which is the protocol the polynomial baseline receives from ridge
cross-validation. In the benchmark harness that fits the spline and
lookup-table families there is no third split, and the family member reported
is the one that scores best on the evaluation set. That choice favors the
baselines rather than the fabric, so it makes every comparison here
conservative in the direction that matters.

On a six-task synthetic suite chosen to include targets that favor an
exponential--logarithmic primitive, we compared a \num{425}-parameter mesh
(depth 4, width 8, ideal hardware) against a \num{433}-parameter MLP and a
\num{20}-parameter polynomial, $n = 3$ seeds, \num{2000} iterations. The
fabric won \num{0} of \num{6}, including on $\exp(x_1/x_2)/x_3$ and
$\exp(x_1) - \exp(x_2)$, the two targets we constructed in its favor. The
fabric-to-MLP test-NRMSE ratios were \num{7.4} (power law), \num{9.9}
(exponential ratio), \num{11.7} (sum of sines), \num{14.6} (triple product)
and \num{20.9} (difference of exponentials), a median of \num{11.7} across
the five targets any model fitted; on $\sin(2\pi x_1x_2)$ all three models
returned NRMSE $\ge 1.0$, a shared failure rather than a result.

On real regression at matched budget the fabric neither wins nor is the
efficient choice. For \texttt{diabetes} the fabric reached test NRMSE
\num{0.7134} at \num{873} parameters against \num{0.7121} for an MLP at
\num{877}; a degree-3 polynomial reached \num{0.6932} at \num{286}. For
\texttt{wine\_proline} the fabric reached \num{0.5726} at \num{1001}
parameters against \num{0.7521} for an MLP at \num{995}, but a
\num{13}-parameter linear model reached \num{0.5695}. Architecture was not the
binding constraint: across six depth/width settings at fixed cell budget,
\texttt{diabetes} test NRMSE moved only over
\numrange{0.7099}{0.7179} on ideal hardware for \numrange{425}{885}
parameters.

\begin{figure}[tp]
\centering
\includegraphics[width=\linewidth]{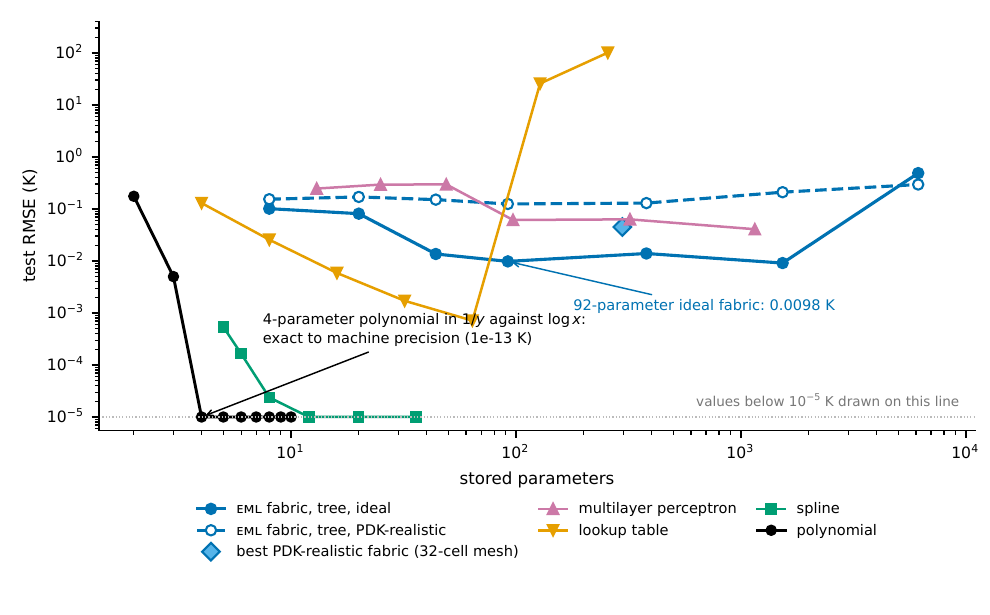}
\caption{\textbf{Accuracy against stored parameters on the sensor-linearization
task.} Test RMSE on a Steinhart--Hart thermistor over
$T \in [-20, 60]\,^{\circ}\mathrm{C}$ against stored parameters, logarithmic
axes. Blue filled circles: the fabric as a binary tree on ideal hardware; blue
open circles: the same trees with PDK-realistic non-idealities; blue diamond:
the best PDK-realistic fabric over all topologies, a 32-cell depth-4 mesh at
\num{297} parameters. Each fabric point is the median of \num{3} seeds, and a
tree of $N$ cells stores $6N+2$ parameters. Baselines are polynomials (black),
splines (green), lookup tables (orange) and multilayer perceptrons (pink), each
at its best input--output transform for that parameter count. The
four-parameter polynomial is the analytic inverse of Steinhart--Hart and is
exact by construction, so the ratio between it and the fabric measures machine
epsilon rather than an accuracy gap; the defensible statement is that the task
admits a four-parameter closed form while the \num{92}-parameter ideal fabric
reaches only \SI{0.010}{\kelvin}. Lookup tables degrade beyond \num{64} entries
because the training set no longer fills their bins.}
\label{fig:accuracy}
\end{figure}

The sharpest loss is on the sensor-linearization task we designed the fabric
for. On a Steinhart--Hart thermistor over
$T \in [-20, 60]\,^{\circ}\mathrm{C}$, a polynomial of degree 3 in $1/y$
against $\log x$ is exact to \SI{0.00000}{\kelvin} at \num{4} parameters, and
a degree-2 form reaches \SI{0.00500}{\kelvin} at \num{3}. The best ideal
fabric on the same target is \SI{0.010}{\kelvin} at \num{92} parameters
(15-cell depth-4 tree), and the best PDK-realistic fabric is
\SI{0.033}{\kelvin} at \num{297}. A four-parameter log-polynomial beats a
\num{92}-parameter ideal fabric on the fabric's own showcase problem. The
transformed polynomials also extrapolate better outside the training range
(\SI{0.217}{\kelvin} against \SI{1285.9}{\kelvin} for the untransformed form).
Figure~\ref{fig:accuracy} places the fabric against every baseline on the same
axes: the ordering does not depend on where the parameter budget is set, and no
fabric configuration at any size reaches the four-parameter polynomial.

Two internal findings shape how the fabric should be built. First, topology
matters and is target-dependent. On the sensor curve (corrected-PDK hardware,
$n = 3$ seeds, \num{3000} iterations, median\,/\,best RMSE in K) the ordering
is mesh $<$ tree $<$ DAG and is not close: mesh d4w8 \num{0.044}\,/\,\num{0.033}
at \num{297} parameters, mesh d4w4 \num{0.089}\,/\,\num{0.046} at \num{149},
tree d4 \num{0.129}\,/\,\num{0.127} at \num{92}, tree d8
\num{0.213}\,/\,\num{0.105} at \num{1532}, DAG d4w4
\num{0.369}\,/\,\num{0.157}, DAG d8w32 \num{10.503}\,/\,\num{8.611} at
\num{15457}. A 32-cell depth-4 mesh beats a 255-cell depth-8 tree by
\num{5}$\times$ on one eighth the cells. On the \texttt{osc\_k3} target the
ordering reverses: the tree gives \numlist{0.313;0.354;0.319} at depths 4, 6
and 8 while mesh and DAG sit at \numrange{0.46}{0.68}.

Second, error does not accumulate with depth in the way a cascade argument
predicts. With frozen weights and no training, the relative RMS divergence
between ideal and mismatched fabrics scales as $d^{\,p}$ from depth 2 upward
with $p = \num{0.35}$ at \SI{1}{\percent} gain sigma, $p = \num{0.38}$ at
\SI{3}{\percent} and $p = \num{1.01}$ at \SI{10}{\percent}. At the corrected
\SI{3}{\percent} the growth is sub-$\sqrt{d}$: going from 7 to \num{1023}
cells costs \SI{57}{\percent} more output error, not \num{12}$\times$. At
\SI{10}{\percent} it is linear and then diverges past depth 8
($\num{4.57e11}$ at depth 10). The instability threshold lies between
\SI{3}{\percent} and \SI{10}{\percent}, and we did not locate it. The apparent
depth-4 turnover reported earlier was an initialization pathology rather than
a property of the fabric: replacing a $\mathcal{N}(0,0.7^2)$ initialization
with a unity-gain one moved the depth-3 ideal RMSE from \num{0.535} to
\num{0.014} and the depth-3 PDK-realistic median from \num{5.337} to
\num{0.162}, a factor of \num{33} at median from initialization alone.

We report the ODE study here for completeness, but it is not resource-matched:
its only comparator is a polynomial of $\le 21$ terms against fabrics of
\numrange{85}{297} parameters, and no MLP baseline exists in it. On
Lotka--Volterra and van der Pol the polynomial is exact by construction, so
the reported ratios measure only that the fabric is not a polynomial. On the
damped pendulum a \num{6}-term degree-5 polynomial is \num{6.15}$\times$
better than a \num{297}-parameter ideal fabric (\num{0.0159} against
\num{0.0978}).

\subsection{Pointwise fit does not predict closed-loop behavior}
\label{sec:res-closedloop}

We substituted trained surrogates for the nonlinear term of \num{16} dynamical
systems and integrated the resulting closed loop. Matching was enforced in
code: the fabric's own parameter count was read back and the MLP hidden width
solved to it, with the same seeds ($n = 3$) and the same \num{3000}
iterations. We under-resourced the polynomial baseline at
\numrange{10}{110} parameters, so that comparison is conservative.
Whether \num{3000} iterations is enough to be fair to the fabric is a
reasonable question, and Section~\ref{sec:res-elm} bounds the answer without
a further sweep: a fabric whose interior weights are never trained at all,
read out the way this one is, performs level with the trained fabric. What
more optimizer effort could buy is therefore bounded by what training buys
over not training, which on this suite is approximately nothing.

On closed-loop trajectory NRMSE the fabric won \num{0} of \num{16}. On
pointwise test NRMSE, over the same runs, it won \num{1} of \num{16}
(\texttt{mm\_bisub}); the median fabric-to-baseline fit ratio was
\num{21.5}$\times$ worse with a range of \numrange{0.92}{3.9e10}. Restricting
the comparison to fabric against MLP alone gives \num{1} of \num{16} on
trajectory (\texttt{logistic}, \num{0.000885} against \num{0.001902}), but the
polynomial beats both. The count is criterion-dependent, and we state the
criterion with every use.

The Michaelis--Menten system is the case where matching inverts the
conclusion. An earlier unmatched comparison put a \num{199}-parameter fabric
against a \num{21}-term polynomial and reported a \num{5.6}$\times$ fabric
advantage. Re-running the same system with MLPs sized to the fabric's own
parameter count gives closed-loop trajectory NRMSE of \num{0.010659} for the
best fabric of any size and \num{0.003531} for the best MLP of any size, a
\num{3.02}$\times$ loss; strictly at matched size (\num{199} against \num{201}
parameters) it is \num{0.019247} against \num{0.003531}, a
\num{5.45}$\times$ loss. A \num{55}-parameter polynomial (\num{0.012915}) also
beats the \num{199}-parameter fabric. The earlier figure is superseded, and we
report the \num{3.02}$\times$ best-of-family form below because it is the most
favorable defensible statement for the fabric.

To separate surrogate quality from the dynamics we quantified error
amplification with no fabric present. We injected a controlled perturbation into the \emph{true} nonlinear term, either a systematic gain error or an unbiased smooth ripple of the same RMS at $\varepsilon \in \{0.01, 0.03\}$, and recorded $A = \text{trajectory NRMSE}/\varepsilon$ over
$n = 60$ runs spanning \num{15} systems. The median $A$ was \num{0.613} with a
range of \numrange{0.046}{19.26}. By structural class the medians were
\num{0.463} for contracting ratio systems ($n = 32$), \num{0.587} for
contracting polynomial systems ($n = 8$), \num{1.492} for the trigonometric
contracting system ($n = 4$), \num{1.588} for oscillatory ratio systems
($n = 12$) and \num{8.721} for oscillatory polynomial systems ($n = 4$);
\num{5} of \num{15} systems had median $A > 1$.

$A$ predicts how far closed-loop error runs ahead of pointwise error, and it
does so for the baselines as well as for the fabric. Regressing
$\log_{10}$ of the trajectory-to-fit NRMSE ratio of each system's
best-fitting model on $\log_{10} A$ over the \num{15} systems gives Pearson
$r = \num{0.917}$ for the fabric family and $r = \num{0.813}$ for the MLP
family. (Including the polynomial, which is exact on several systems and makes
the ratio numerically ill-conditioned, drops the baseline correlation to
$r = \num{0.400}$.) The median trajectory-to-fit degradation was \num{1.14}
for the fabric and \num{1.55} for the baselines. Closed-loop degradation is
therefore a property of the system being closed, not a fabric-specific pathology. The fabric, however, enters that regime from a worse pointwise fit.

Conservation is destroyed at pointwise errors that look tolerable.
Lotka--Volterra carries a conserved quantity, and we record its relative drift
over the integration alongside the fit. At a test NRMSE of \num{0.030}, the
\num{85}-parameter pedestal fabric drifts the invariant by \num{0.436} and
reaches trajectory NRMSE \num{0.409}; the \num{199}-parameter ideal fabric
fits to \num{0.054} and drifts by \num{6.278}. The matched MLPs fit to
\num{0.006} and still drift by \num{0.276} (\num{85} parameters) and
\num{0.513} (\num{201} parameters). Only the \num{6}-parameter polynomial,
exact by construction, holds the invariant (\num{7.8e-6}). A three-percent
pointwise error coexists with a forty-percent loss of the invariant, in the
fabric and in its matched baseline alike.

\subsection{What the extracted layout licenses}
\label{sec:res-silicon}

Every device parameter in this paper is extracted from one layout, and what
that layout licenses is narrower than what it contains. Geometric and connectivity
checking is clean under the conditions given in Methods, which is what permits
reading the extracted netlist as an electrical description of the layout. It permits nothing more than that. The tile it sits on is described in
Supplementary Note~S4; its area divides in the same proportion
Table~\ref{tab:blockbudget} gives for energy, the two cells taking a third and
the bias core, weight DACs and scan chain that make them usable taking the
rest.

We extracted the shipped cell with resistance and coupling (\num{61} devices,
\num{4984} parasitic resistors, \num{1316} capacitors). Its DC transfer moves
by less than \SI{0.25}{\percent} against a parasitic-free control; under a
\SI{500}{\nano\ampere}, \SI{20}{\nano\second} current kick none of \num{25}
corner and temperature conditions sustained an excursion. Post-layout, one
cell draws \SI{29.87}{\micro\ampere} from \SI{3.3}{\volt}
(\SI{98.57}{\micro\watt}) and settles to \SI{1}{\percent} in
\SI{402}{\nano\second} and to \SI{0.1}{\percent} in \SI{497}{\nano\second}
with \SI{2.17}{\percent} overshoot. We extracted the two-cell assembly in one
piece (\num{410} devices, \num{10971} resistors, \num{3174} capacitors), and it
settled at all \num{8} conditions that integrated; the ninth, \texttt{ss} at
\SI{-40}{\celsius}, did not integrate and is reported as unavailable. Nothing
deeper than two cells was extracted.

The transfer functions are
$i_o = \num{1.101}\exp(\num{0.967}u) - \num{0.009}$ (maximum residual
\num{4.3e-4} units) and
$i_o = \num{2.478} - \num{1.094}\ln(v + \num{2.547})$ (maximum residual
\num{2.2e-4} units). A joint two-port sweep (\num{102} of \num{102} operating
points converged) fits the separable form $A\exp(au) + k\ln(v+s) + c$ to a
maximum residual of \num{0.0034} units over a \num{5.354}-unit output span,
with a multiplicative cross term of $-\num{0.0000}$: within that resolution
the cell is the sum of its two paths. The submitted part composes through the $v$ port only. The $u$ swing at every downstream port is \num{0.000000} units in every weight configuration built, so it computes
$-\ln(-\ln x)$, and every exponential-port composition quantity in this paper
is derived from the extracted exponential fit and the simulated device span
rather than swept end to end. These fits, the \num{440}-sample mismatch Monte
Carlo and the four-cell chain sweep are the whole of the silicon's
contribution to this paper.

\subsection{Silicon enters the software model as parameters, and bounds its depth}
\label{sec:res-nonideal}

The silicon enters the software study as parameters. DC sweeps of the netlist
that LVS-matches the submitted layout give
$i_o = \num{2.478} - \num{1.094}\ln(v + \num{2.547})$ on the segmented
resistor basis, so the cell computes $\ln(v+s)$ with a hard-wired pedestal
$s = \num{2.547}$ units; the lumped LVS basis gives $s = \num{2.574}$ and
slope \num{1.145}, \SI{1.05}{\percent} and \SI{4.5}{\percent} away
respectively. All studies below use the segmented basis. The extracted device
span is \num{8.372} decades (\numrange{5.80e-6}{1366} units).

These two effects are not independent, and combining them double-counts the
same physics. The per-hop $v$-port gain
$\lambda_v = \num{1.094}/(\num{1.741} + \num{2.547}) = \num{0.2551}$ is
\emph{derived} from the pedestal. A probe of the realized per-hop gain
confirms it: \num{0.6284} ideal, \num{0.2551} with the pedestal alone, which reproduces the extracted silicon value exactly, \num{0.6284} with the attenuation alone, and \num{0.0933} with both, \num{2.7}$\times$ too small.
Every study reported here uses the pedestal alone. The device span does not
bind: at \num{8.372} decades the over-excursion is \num{0.00} decades in every
workload tested, and only an artificial \num{2}-decade stress configuration
produces a real over-excursion (\num{2.171} decades).

Injecting the pedestal into the software model does not degrade the fit. Over
\num{32} paired ideal/pedestal comparisons in the closed-loop study
(\num{16} systems $\times$ 2 cell counts), the median pedestal-to-ideal test
NRMSE ratio was \num{0.852} (range \numrange{0.028}{3.758}) and the pedestal
fabric was the better of the pair in \num{19} of \num{32}; on trajectory the
median ratio was \num{0.969} and the pedestal was better in \num{18} of
\num{32}. A hand-programmed $2\times2$ DAG computing $x_1x_2$ makes the
mechanism explicit: its relative RMS is \num{2.023} under the pedestal and
\num{0.056} after refitting the output stage, whereas under the $v$-port
attenuation alone it goes from \num{0.821} to \num{4.74e-16}. Retraining
absorbs the attenuation fully; the pedestal is absorbed to within a factor
of \num{36} but not to machine precision.

The calibration route dominates every individual device effect. With all
non-idealities enabled at once (mismatch \SI{5}{\percent}, per-stage noise
\num{3e-3}, saturation 30, 8-bit weights) on a depth-3 tree, $n = 8$ seeds,
median\,/\,best RMSE in K: in-situ training gives
\num{6.3}\,/\,\num{0.94} on the $\beta$ model and \num{3.6}\,/\,\num{0.76} on
Steinhart--Hart, while transferring factory-trained weights gives
\num{27.8}\,/\,\num{15.0} and \num{25.0}\,/\,\num{5.5}, a factor of \numrange{4.4}{7.7} at median. On a PDK-calibrated chip, in-situ training
gives \num{5.5}\,/\,\num{0.77} and \num{3.8}\,/\,\num{0.86}; an uncompensated
$\pm\SI{10}{\percent}$ drift gives \num{30.6}\,/\,\num{4.4}, a
$\pm\SI{1.5}{\percent}$ PTAT residual gives \num{9.7}\,/\,\num{1.1}, and the
same $\pm\SI{10}{\percent}$ drift with a 300-iteration retrain recovers to
\num{14.9}\,/\,\num{4.0}. Taken one at a time on the same depth-3 tree
($n = 6$ seeds, median\,/\,best in K, ideal \num{0.83}\,/\,\num{0.18}),
saturation at 3 is worst (\num{10.6}\,/\,\num{0.35}), followed by 4-bit
weights (\num{7.07}\,/\,\num{4.29}), 8-bit weights
(\num{2.25}\,/\,\num{1.46}) and \SI{5}{\percent} mismatch
(\num{2.08}\,/\,\num{0.36}); mild noise at \num{3e-3} improved the best seed
over ideal (\num{0.08} against \SI{0.18}{\kelvin}).

One extracted quantity bounds which fabric configurations can be modeled at
all. A \num{440}-sample post-layout Monte Carlo with device mismatch enabled
gives a level sigma of \num{0.266} units against \num{1.288} units of swing,
\num{2.28} bits. Circuit simulation of the fabricated chain gives a
signal-to-sigma ratio of \num{4.84} at one cell, \num{0.775} at two,
\num{0.214} at three and \num{0.0517} at four. Uncalibrated $v$-direction
configurations deeper than one cell are therefore below the shipped part's own
mismatch floor, and we report no projection at those depths.

\subsection{Scaffolding, not the operator, sets the energy}
\label{sec:res-energy}

A single evaluation occupies one \SI{402}{\nano\second} settling
window. Inside that window the four bipolar junctions that compute
$\mathrm{eml}$ move \SI{13.0}{\femto\joule} onto the
\SI{47.9}{\femto\farad} of their own extracted capacitance, and draw
nothing from the \SI{3.3}{\volt} rail at all, because their current
arrives through the pedestal legs at \SI{1.2}{\volt}.  Over the same
window the four servo amplifiers spend \SI{29.53}{\pico\joule} holding
the junctions' collectors where the reference put them. They compute
nothing. Neither does the bias network, which spends
\SI{8.70}{\pico\joule} supplying operating currents, nor the output
mirrors, which spend \SI{1.39}{\pico\joule} copying the answer
onward. One evaluation therefore costs \SI{39.6}{\pico\joule}, of
which the arithmetic is \SI{13.0}{\femto\joule}. Everything below is a
way of taking that sentence seriously.

The energy law puts the same statement in a form that transfers. Substituting the cell's own extracted quantities into \eqref{eq:eop} gives $\SI{29.3}{\pico\joule}$ with nothing fitted. Those quantities are $F = \num{15.1}$, $N_\tau = 402/45.9 = \num{8.76}$, $C = \SI{2.00}{\pico\farad}$, $V_\mathrm{dd} = \SI{3.3}{\volt}$ and $nV_T = \SI{33.6}{\milli\volt}$. Two of those five terms
belong to the scaffolding and none belongs to the operator: $F$ counts how
many times the computing branch's current is drawn from the supply, and $C$ is
the capacitance that has to be charged to keep the servo holding it stable.
Driving both terms to their floors, $F = 1$ with $C$ at parasitics, divides that estimate by about \num{600} and brings it to a few tens of
femtojoules, the order of the core itself, though the law is least accurate
at that end and overshoots the directly priced \SI{13.0}{\femto\joule} by
\num{3.7}. That floor is also unreachable, since $F = 1$ means the collector
voltages float and the translinear law fails. The reachable version of the same move is sharing the scaffolding across cells rather than abolishing it. That is worth \num{144} and is priced in the Discussion. The distance between
that factor and the \num{3046} above is the subject of this paper.

The energy law \eqref{eq:eop}, derived in Methods, predicts the cell forward
from extracted inputs with nothing fitted, and lands within \SI{26}{\percent}
on both axes: $\tau$ predicted \SI{33.9}{\nano\second} against
\SI{45.9}{\nano\second} extracted, and $E$ predicted \SI{29.3}{\pico\joule}
against \SI{39.6}{\pico\joule}, both ratios \num{0.74}. Its two variable terms
are $F = \num{15.1}$ and $C = \SI{2.00}{\pico\farad}$, and the bias current
cancels out of it exactly.

The assumption-free result is the composability overhead. Price the bare translinear core, \num{4} bipolar junctions and \num{4} poly resistors carrying \SI{3.962}{\micro\ampere} at \SI{1.2}{\volt}, with
$C_{\text{core}} = \SI{47.9}{\femto\farad}$ taken pessimistically as the
core's \SI{2.7}{\percent} area share of the extracted capacitance,
$g_m = \SI{153.3}{\micro\siemens}$, $\tau = \SI{0.31}{\nano\second}$ and
$t_{\text{settle}} = \SI{2.7}{\nano\second}$, and it gives $E_{\text{core}} = \SI{13.0}{\femto\joule}$ against
\SI{39.6}{\pico\joule} for the cell as built. The ratio is
\textbf{\num{3046}$\times$}, and it is a per-cell quantity: seven cells pay
it seven times over rather than diluting it, which is the whole of the
composability claim. It is $E_{\text{fab}}/E_{\text{core}}$ and contains no
digital model, so it is constant across the \num{72} corners of the
sensitivity box described below by construction, not by good fortune. The
corresponding seven-cell figure is \num{1541}$\times$ (\SI{420.8}{\pico\joule} against
\SI{0.273}{\pico\joule}), with both terms charged the full three-stage chain
latency. We report the per-cell ratio as the
headline because it is free of that latency convention, both of its terms
being one device settling once.
The corresponding ratios are \num{16}$\times$ in area and
\num{5.6}$\times$ in supply current, so energy is the harshest of the three.
Table~\ref{tab:blockbudget} resolves the cell block by block, and Supplementary
Note~S5 walks the accounting through; together they answer the
question the \SI{2.7}{\percent} figure raises: if the junctions require almost
nothing, what does. The blocks that evaluate the operator draw
\SI{0}{\percent} of the rail current, while the four servo amplifiers that
exist only to hold their operating points take \SI{42.5}{\percent} of the area
and \SI{74.5}{\percent} of the supply current and \emph{are} $F = \num{15.1}$;
pedestal legs and output mirrors account for the rest of the static draw, and
a further quarter of the area scales the inputs and stores the codes that do
so. A coupling link costs \SI{88}{\percent} of a cell in power against
\SI{28}{\percent} in area. Both terms in \eqref{eq:eop} are scaffolding
terms, and the dominant cost of an $\mathrm{eml}$ cell is not the
transcendental but holding the operating points that make the transcendental
accurate.

\begin{table}[tb]
\centering
\footnotesize
\setlength{\tabcolsep}{3pt}
\caption{\textbf{Per-block area, static current and energy for one
$\mathrm{eml}$ cell.} Area is the extracted per-cell marginal silicon of the shipped GDS: the \SI{10874}{\micro\meter\squared} analog cell block plus the
two input weight DACs and their \num{10} configuration flops,
\SI{15016.5}{\micro\meter\squared} in total, which closes against the placed
two-cell floorplan to \SI{0.1}{\micro\meter\squared}. The
\SI{2.7}{\percent} core share quoted above is \num{297}/\num{10874}, the core
against the analog cell block alone. Current is the extracted operating point
of the isolated cell in die context, \SI{29.8686}{\micro\ampere} from
\SI{3.3}{\volt} $=\SI{98.566}{\micro\watt}$; the fitted in-chain slope
\SI{96.95}{\micro\watt} is not used here. Energy is each block's share of
$\SI{98.566}{\micro\watt} \times \SI{402}{\nano\second} =
\SI{39.62}{\pico\joule}$, the settling window being common to all blocks.
Weight-DAC supply current was metered at code \num{0} only, so that row, and
therefore every energy figure in this table including the total, is a lower
bound.}
\label{tab:blockbudget}
\begin{tabular}{@{}lrrrrrp{0.22\linewidth}@{}}
\toprule
Block & Area & \% & $I_{vdd}$ & \% & $E$ & What it is for \\
 & (\si{\micro\meter\squared}) & & (\si{\micro\ampere}) & & (\si{\pico\joule}) & \\
\midrule
Translinear core (\num{4} NPN)      & 297    & 1.98  & 0        & 0    & 0     & evaluates $\mathrm{eml}$ \\
\midrule[\cmidrulewidth]
Exponent-scale resistors (\num{4} poly)\textsuperscript{a}
                                    & ---    & ---   & 0        & 0    & 0     & sets the exponent scale \\
\midrule[\cmidrulewidth]
Servo amplifiers (\num{4})          & 4621   & 30.77 & 22.2587  & 74.5 & 29.53 & holds the junction nodes at their set points \\
\midrule[\cmidrulewidth]
Compensation capacitance (\num{4} MiM)
                                    & 2296   & 15.29 & 0        & 0    & 0     & keeps those servo loops stable \\
\midrule[\cmidrulewidth]
Bias network (pedestal legs, FETs, resistors)\textsuperscript{a}
                                    & 1444   & 9.62  & 6.5605   & 22.0 & 8.70  & supplies the core's operating currents \\
\midrule[\cmidrulewidth]
Output mirrors (\num{2} PFET)\textsuperscript{b}
                                    & ---    & ---   & 1.0494   & 3.5  & 1.39  & copies the result to the next cell \\
\midrule[\cmidrulewidth]
Weight DACs (\num{2} $\alpha$, \num{4}\,b${+}$sign)
                                    & 2838.6 & 18.90 & $<0.0001$\textsuperscript{c} & 0.0 & $\geq 0$\textsuperscript{c} & scales the two inputs \\
\midrule[\cmidrulewidth]
Weight configuration flops (\num{10})
                                    & 1304.0 & 8.68  & ---\textsuperscript{d} & --- & ---\textsuperscript{d} & holds the weight code between writes \\
\midrule[\cmidrulewidth]
In-cell routing, fill and tap strips
                                    & 2216   & 14.76 & 0        & 0    & 0     & connects the blocks \\
\midrule
Total, one cell                     & 15016.5\textsuperscript{e} & 100.00 & 29.8686 & 100.0 & 39.62 & \\
\bottomrule
\end{tabular}

\vspace{2pt}
{\footnotesize
\textsuperscript{a}\,The four exponent-scale poly resistors are not separable
in the layout record: their area is counted together with the bias FETs in the
\SI{1444}{\micro\meter\squared} row. Neither draws current from
\texttt{vdd}; the resistors carry the core's \SI{3.962}{\micro\ampere} of
collector current, which the pedestal legs supply.
\textsuperscript{b}\,The output mirrors are inside the analog cell block but
are not broken out as a separate area in the layout record; their area is
distributed over the bias and routing rows.
\textsuperscript{c}\,\SI{<0.1}{\nano\ampere} total across the four DACs of the
shipped chain, at code \num{0} only. Non-zero codes were never simulated.
\textsuperscript{d}\,Flop switching energy is not metered; the flops are static
between weight writes.
\textsuperscript{e}\,The nine component areas are recorded to
\SI{1}{\micro\meter\squared} and sum to \SI{15016.6}{\micro\meter\squared},
\SI{0.1}{\micro\meter\squared} above the closure-checked total.
\par}
\end{table}

\begin{table}[tb]
\caption{\textbf{The sensitivity box: what is assumed, over what range, and
on what authority.} The \num{72} corners are the product of these three
axes. The composability overhead contains none of them, being a ratio of two extracted analog quantities, which is why it is invariant across the box by construction rather than by robustness.}
\label{tab:sensitivity}
\centering\small
\begin{tabular}{@{}p{0.20\linewidth}p{0.20\linewidth}p{0.50\linewidth}@{}}
\toprule
Axis & Range (central) & Provenance \\
\midrule
Fabric precision
  & \numrange{2.28}{5.73} bits (anchor \num{3.86})
  & anchor is the \textsc{pdk}-realistic in-situ median of \num{8} seeds,
    \SI{5.5}{\kelvin} over an \SI{80}{\kelvin} span; the band runs from the
    \num{440}-sample post-layout mismatch floor to the slope-mismatch
    ceiling \\
\midrule[\cmidrulewidth]
Control-overhead multiplier
  & \numrange{2}{10} (\num{3.0})
  & applied to the digital energy-per-operation model; assumed, not
    extracted, which is why it is swept \\
\midrule[\cmidrulewidth]
Node-scaling factor
  & \numrange{5.14}{11.56}
  & \SI{45}{\nano\meter} to \SI{130}{\nano\meter}, carrying Horowitz's
    per-operation figures~\cite{horowitz2014computing} to this process \\
\bottomrule
\end{tabular}
\end{table}

\begin{figure}[tp]
\centering
\includegraphics[width=\linewidth]{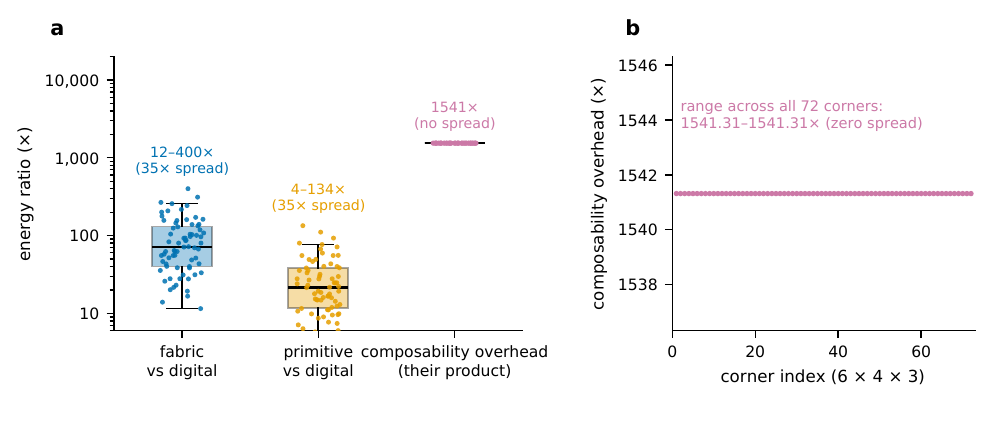}
\caption{\textbf{Both comparisons with digital hardware move by a factor of
\num{35} across the study's assumptions; the ratio between them does not move at
all.} Each point is one of the \num{72} corners of the sensitivity box:
\num{6} values of the precision assigned to the fabric
(\numrange{2.28}{6.70} bits) $\times$ \num{4} values of the digital
control-overhead multiplier (\numrange{2}{10}) $\times$ \num{3} values of the
\SI{45}{\nano\metre}-to-\SI{130}{\nano\metre} scaling factor
(\numrange{5.14}{11.56}).
\textbf{a}, Distribution over those corners of the three headline ratios. The
seven-cell fabric loses to a width-matched digital datapath by
\numrange{12}{400}$\times$, and the bare translinear primitive beats the same
datapath by \numrange{4}{134}$\times$; both spans are about \num{35}$\times$
wide and both depend entirely on a model of digital energy. The digital model
cancels identically from their product, which is the seven-cell
$E_{\text{fabric}}/E_{\text{core}}$ and is \num{1541}$\times$ at every corner.
Both terms of that product charge all seven cells, and all seven bare cores,
the same three-stage chain latency. The paper's headline quantity is the
per-cell ratio, \num{3046}$\times$, which is free of any latency convention. Boxes give the
quartiles with whiskers at \num{1.5} interquartile ranges; the individual
corners are overplotted.
\textbf{b}, The same product against corner index, on a vertical axis fine
enough to resolve one part in a thousand. It is a straight line: the two
conclusions that depend on external assumptions never change sign anywhere in
the box, and the one that does not depend on them never changes value.}
\label{fig:sens}
\end{figure}

Comparisons against digital depend on two modeling constants and on the
precision assigned to the fabric, so we report them as ranges over the full
box. Table~\ref{tab:sensitivity} gives the three axes, their ranges and their
provenance. We anchor the fabric at \num{3.86} bits, the PDK-realistic in-situ median of \num{8} seeds at \SI{5.5}{\kelvin} over an \SI{80}{\kelvin} span, with a reachable band of \numrange{2.28}{5.73} bits, and sweep the control
overhead multiplier over its defensible range (\numrange{2}{10}, central
\num{3.0}) and the \SI{45}{\nano\meter}-to-\SI{130}{\nano\meter} scaling
factor over \numrange{5.14}{11.56}. Across all \num{72} corners, the
seven-cell fabric loses to a width-matched digital datapath by
\numrange{12}{400}$\times$ (worst case \num{11.5}$\times$, at \num{6.70} bits
with the largest control overhead), and the bare primitive beats the same
datapath by \numrange{4}{134}$\times$. Neither conclusion flips anywhere in
the box, and the digital model cancels identically from their product, which
is why the composability overhead carries no digital assumption. That
cancellation is an algebraic identity and the flat line it produces is a
check that the implementation honors it, not independent evidence of
robustness. The
previously published point values of \num{30}$\times$ and \num{152}$\times$
are one corner of this box, obtained at control overhead \num{3.0}, scaling
factor \num{11.56} and a fixed 8-bit baseline. That corner is not a central
case and we do not present it as one: \num{3.0} sits at the bottom of the
defensible control-overhead range, which is the least favorable choice for the
fabric, while \num{11.56} sits at the top of the node-scaling range, which is
the most favorable. The two errors point in opposite directions and the range,
not the corner, is what should be cited. Further, the \num{152}$\times$
additionally charged seven bare cores one settling time each rather than the
chain latency the fabric is charged, and at that same corner the
chain-consistent value is \num{51}$\times$. We do not use the value
\num{4.30} bits that appeared in earlier work: it traces to the single best of
the \num{9} ideal-hardware runs on one \num{9}-variable regression over \num{3} seeds and \num{3} variable-selection modes, at NRMSE \num{0.0507}, whose remaining \num{8} runs give \numrange{2.39}{3.76} bits,
median \num{2.91}.
Figure~\ref{fig:sens} shows the whole box. The two assumption-dependent ratios
each move by about \num{35}$\times$ across it while neither changes sign, and
their product is flat to one part in a thousand: the composability overhead is the
only one of the three quantities that a reader need not accept a digital energy
model to accept.

\begin{figure}[tp]
\centering
\includegraphics[width=\linewidth]{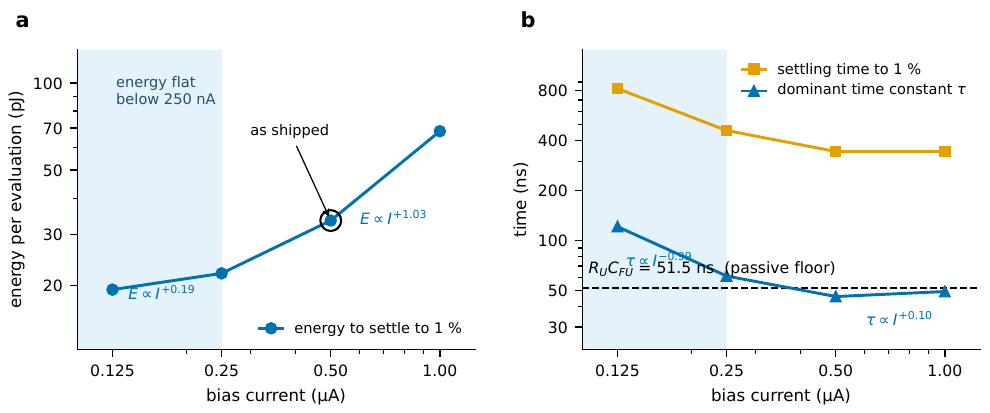}
\caption{\textbf{Energy per evaluation is nearly independent of bias current
below a knee, and the settling floor is set by a passive time constant.}
\textbf{a}, Energy to settle one cell to \SI{1}{\percent} against bias current
over an 8:1 range. Below \SI{250}{\nano\ampere} (shaded) the bias current
cancels out of $E = P\,t$ almost exactly, $E \propto I^{+0.19}$, and the energy
moves only from \SI{22.00}{\pico\joule} to \SI{19.33}{\pico\joule}; above
\SI{500}{\nano\ampere} the cancellation breaks, $E \propto I^{+1.03}$, and the
extra current is spent for nothing. The circled point is the bias as submitted,
on the far side of the knee. \textbf{b}, The reason. The dominant time constant
falls as $I^{-0.99}$ below the knee, as $C/g_m$ requires, then flattens once it
meets a fixed pole in the exponential path: \SI{51.5}{\kilo\ohm} into
\SI{1}{\pico\farad} gives \SI{51.5}{\nano\second} (dashed), against a $\tau$
flooring at \SIrange{45.9}{49.2}{\nano\second}. The floor is a resistor, not a
transistor, and no sizing removes it. There is therefore no bias current at
which this cell is cheap: lowering it stops near \SI{19}{\pico\joule} for
\num{2.4}$\times$ the latency, and raising it is wasted. Four bias points from
an \texttt{ngspice} sweep in die context, schematic basis.}
\label{fig:bias}
\end{figure}

A direct ngspice bias sweep (sky130A, \texttt{tt}, \SI{27}{\celsius},
\texttt{gear}/\texttt{maxord=2}, \SI{2}{\nano\second} grid, same die context
and normalized operating point at four bias points over an 8:1 range,
schematic basis) tests the cancellation in \eqref{eq:eop} directly:
\SI{0.125}{\micro\ampere} gives \SI{19.33}{\pico\joule} at
\SI{821.0}{\nano\second}, \SI{0.250}{\micro\ampere} gives
\SI{22.00}{\pico\joule} at \SI{458.9}{\nano\second},
\SI{0.500}{\micro\ampere} (as shipped) gives \SI{33.49}{\pico\joule} at
\SI{342.9}{\nano\second}, and \SI{1.000}{\micro\ampere} gives
\SI{68.17}{\pico\joule} at \SI{342.8}{\nano\second}. Below
\SI{250}{\nano\ampere} the cancellation holds,
$\tau \propto I^{-0.99}$ and $E \propto I^{+0.19}$; above
\SI{500}{\nano\ampere} it breaks, $\tau \propto I^{+0.10}$ and
$E \propto I^{+1.03}$, so extra current buys no speed. The knee is set by a
resistor rather than a transistor: the exponential path runs through
$R_U$ (\texttt{res\_high\_po}, \SI{51.5}{\kilo\ohm}) into
$C_{FU} = \SI{1}{\pico\farad}$, and $R_U C_{FU} = \SI{51.5}{\nano\second}$
against a simulated $\tau$ that floors at
\SIrange{45.9}{49.2}{\nano\second}. Lowering the bias floors the cell at about
\SI{19}{\pico\joule} (\num{1.73}$\times$ cheaper for \num{2.4}$\times$ the
latency); raising it is wasted. There is no bias current at which this fabric
is cheap. Figure~\ref{fig:bias} shows both halves of the measurement: the
energy plateau below the knee in panel~\textbf{a} and, in panel~\textbf{b}, the
$\tau$ floor that causes it, sitting on the passive $R_UC_{FU}$ line.

We tested two further limits. Thermal $kT/C$ noise does not bind: at
\numrange{2.28}{5.73} bits it permits capacitances of order
\numrange{1e-19}{1e-18}\,F, so the \SI{2}{\pico\farad} present is about
\num{1e6} larger than noise requires, and $kT/C$ does not bind until
\num{14.4} bits at a \SI{1}{\volt} swing. What sets $C$ is loop stability, and
what sets precision is matching. The correct fundamental floor for a
current-mode translinear circuit is shot noise on the delivered charge,
$E_{\min} = V_{\text{swing}}\,q\,4^{b}$, which crosses the
\SI{130}{\nano\meter} digital datapath at $b^\ast = \num{13.8}$ bits: the
fabric operates \numrange{8}{11} bits inside the regime where analog is
fundamentally cheaper and loses anyway, spending the charge of a
\num{13.9}-bit computation to deliver its own few bits.

Of six candidate application regimes, five fail on the numbers as published.
Always-on monitoring fails structurally, since the break-even rate
(\SI{7.1}{\mega\hertz} for one cell, \SI{86.4}{\mega\hertz} at $N = 7$)
exceeds the settling-limited ceiling ($\SI{2.49}{\mega\hertz}$ and
$\SI{2.84}{\mega\hertz}$). Analog-in/analog-out fails because it would need
\numrange{187}{748} conversions per evaluation at a fair
\numrange{4}{6}-bit converter against at most \num{14} available ports. The transcendental regime is the exception: at the
published corner the fabric loses to a CORDIC evaluation by \num{1.11}$\times$,
but across the \num{72}-corner box it \emph{wins} that comparison in \num{27}
corners. The regime nonetheless fails against the baseline a designer would build. A 16-entry exponential lookup table with interpolation costs \SIrange{1.44}{16.17}{\pico\joule} across the box, and the fabric loses
by \numrange{2.5}{28}$\times$ everywhere.

All fabric energy figures above are lower bounds: we never simulated
weight-DAC energy at non-zero codes. The omission can be bounded even
though it was not metered. Each port carries a radix-4 signed current
\textsc{mdac} of full scale $\pm\num{3.75}$ units, and one unit is
\SI{0.5}{\micro\ampere}, so two ports drawing full scale simultaneously add
at most \SI{3.75}{\micro\ampere} to the cell's
\SI{29.87}{\micro\ampere}. That worst case, which no trained configuration
would hold on both ports at once, raises the cell to
\SI{44.6}{\pico\joule}, the composability overhead to about \num{3400} and
the fabric-versus-digital range to \numrange{13}{450}$\times$. Every one of
those moves against the fabric, so the omission cannot reverse a comparison
reported here; it can only make the reported figures conservative by up to
\SI{13}{\percent}. Two per-cell power figures are on record
and we use each in its own context: \SI{98.566}{\micro\watt} for the isolated
cell in die context (the basis of the \SI{39.6}{\pico\joule} per cell, of
Table~\ref{tab:blockbudget} and of the composability overhead) and
\SI{96.95}{\micro\watt} as the fitted per-cell
slope of the in-chain power law
$P(N) = \num{96.95}N + \num{85.64}(N-1) + \num{2.09}$\,\si{\micro\watt} (the
basis of the \SI{420.8}{\pico\joule} seven-cell figure). The two differ by
\SI{1.7}{\percent} because the in-chain slope absorbs the shared bias
distribution that an isolated cell carries alone, and both trace to the same
power report.

\subsection{Non-idealities are free when interior values need only be repeatable}
\label{sec:res-elm}

Every comparison above trains all of the fabric's weights and requires each
interior cell to hold a particular value. That requirement is what the
scaffolding of Section~\ref{sec:res-energy} is bought to satisfy, so the
sharpest test of the diagnosis is to remove it. We draw the interior weights
once, freeze them, treat every cell output as a feature and fit only a linear
readout, by ridge regression with the penalty chosen by generalized
cross-validation. No interior node then has a target value: a cell has only to
be repeatable, not accurate.
The difference is worth making concrete. In the trained fabric a given
interior cell has a job: the fitted expression requires that cell to produce a
particular value for a particular input, and a cell that produces something
else by a few percent has corrupted the expression the cells downstream are
evaluating. In the frozen fabric the same cell has no such job. Whatever it
produces is what the readout was fitted against, so the only thing that can
go wrong is producing something \emph{different} on a later pass. The first
requirement is accuracy and is charged for in servo current; the second is
repeatability and is nearly free, because a cell that is merely consistent
with itself needs no external reference to be held against. This is an extreme learning machine rather than a
reservoir, the fabric being feedforward with no fading memory, and the
recurrence in the closed-loop column belongs to the differential equation and
not to the device. Protocol, seeds and per-system tables are in Supplementary
Note~S12; everything not named here is
Section~\ref{sec:res-matched}'s, unchanged, so the rows are comparable.

The effect on the non-idealities is the result. Adding the extracted mismatch,
the output rail and \num{8}-bit weights on top of the pedestal costs the
trained fabric a factor of \num{2.56} on pointwise fit and \num{1.92} in
closed loop, and it is the better configuration on \num{1} of the \num{16}
systems. The same three effects applied to the frozen fabric cost \num{0.95} and \num{0.99}, which is nothing and marginally favorable, and are the better configuration on \num{12} of \num{16}. The non-idealities are expensive
exactly when interior values have to mean something and free when they do not.
Nothing in that comparison involves an energy model, a digital baseline or a
node-scaling assumption; it is the composability argument arriving from the
behavioral side.

The regime does not, however, rescue the fabric. Frozen and carrying the full
silicon configuration it reaches a median NRMSE of \num{0.0073} on the box and
\num{0.0054} in closed loop, against \num{0.0745} and \num{0.0889} trained, so
freezing is worth an order of magnitude; but the resource-matched multilayer
perceptron reaches \num{0.0054} and \num{0.0034} and the polynomial \num{0.0069}
and \num{0.0032}. Scored by the criterion of
Section~\ref{sec:res-matched}, the frozen fabric wins \num{1} of \num{16} on
pointwise fit, unchanged, and \num{4} of \num{16} on closed-loop trajectory
against the trained fabric's zero. The gap narrows and does not close.

Two further readings bound what the improvement means. The first is that the
comparison against a textbook random-feature machine is lost outright: a
random Gaussian projection through $\tanh$, with the same feature count and
the same readout, beats the fabric by a median \num{3.26}$\times$ on the box
and \num{8.75}$\times$ in closed loop. Whatever the operator's grammar is
worth, it is not a better random projection than the cheapest nonlinearity
available. The second is that the order of magnitude freezing buys is not
bought by freezing. A frozen fabric read out at layer~0 only, which is the
readout the trained fabric uses, is level with the trained fabric, at median ratios \numrange{0.82}{1.95} across the four matched pairs, winning \num{4} to
\num{10} of \num{16}. The improvement comes from reading every cell rather
than the output layer alone, and an all-cell readout requires an output path
per cell rather than per output cell, which is another per-cell peripheral of the kind Section~\ref{sec:res-energy} prices.
Table~\ref{tab:blockbudget} prices that path: the two output mirrors already
in the cell cost \SI{1.39}{\pico\joule}, \SI{3.5}{\percent} of the cell.
Duplicating them so that every cell reaches a readout bus rather than only
its successor therefore costs of order \SI{1.4}{\pico\joule} per cell,
taking the cell to about \SI{41}{\pico\joule} and the composability
overhead from \num{3046} to roughly \num{3150}. Against the order of
magnitude in accuracy that the wider readout buys, the analog copy is cheap,
and it does not erase the gain. Two costs are not in that figure and should
be stated with it: the conversion at the end of the bus, which is shared
across cells and amortizes, and the area and routing of the bus itself, which does not. The mirrors have no separately attributable area in our layout
record, so we can price this readout in energy and not in area. Gradient descent
through the behavioral model is not what limits the trained fabric, and the
median \num{21.5}$\times$ of Section~\ref{sec:res-matched} is not an
optimization artifact.

\section{Discussion}
\label{sec:discussion}

The design premise was that a universal primitive licenses a uniform machine.
The premise survives at the level of the primitive and fails at the level of
the machine, and the two halves are separated by a single simulated ratio.
Bare, the translinear core is exactly as inexpensive as the completeness
result implies; the cell that ships around it costs \num{3046} times as much,
and three of the four blocks that spend the difference evaluate nothing. We
name that quotient the \emph{composability overhead}. Its most useful
property is what it does \emph{not} contain: both terms are analog quantities
from one post-layout netlist, so no energy-per-operation model, node-scaling
factor or assumed precision appears in either, and the ratio is therefore
invariant across this study's assumption space by construction rather than as
a finding. The comparisons that do involve a digital model are weaker and
belong in the ranges Section~\ref{sec:res-energy} quotes.

An overhead of this shape is not peculiar to this fabric, and
Table~\ref{tab:fabrics} sets the precedent out: a cheap computing element
beside expensive peripheral machinery recurs across crossbar in-memory
computing, a continuous-time hybrid machine and an analog
Kolmogorov--Arnold network, in different primitives, topologies and
processes~\cite{shafiee2016isaac,murmann2021mixedsignal,haensch2019next,guo2016hybrid,escudero2026akan}.
The reading transfers to analog neuromorphic hardware, which is also a set of
nonlinear nodes joined by programmable weighted interconnect: the question we
put to $\mathrm{eml}$ can be put to a silicon neuron, and where the state is a
voltage on a capacitor the constant differs but the structure does not,
because a value that is not an attractor still needs something external to
hold it. Physics-based neuromorphic proposals are usually argued on the
cheapness of the device that computes~\cite{markovic2020physics}, which our
results say is already free. What separates this fabric from the crossbar is
not that a tax exists but that it never amortizes: a crossbar divides its
converters across a column, so the cost per operation falls as the array
grows, while a fabric whose cells must each be held divides by nothing. That
is why topologies that share scaffolding are the direction to pursue, and
Section~\ref{sec:res-matched} shows it in miniature.

Whether a better cell escapes it is answered by the structure of the energy
law rather than by any layout. Three of its five terms are fixed by the
process, by device physics and by the precision the task demands; all of a
designer's freedom sits in $F$ and $C$, and the translinear principle stops
either reaching its floor. The law holds only while each junction's collector
is held against the Early effect, holding a node requires an amplifier that
draws current, and an amplifier in feedback requires compensation. A
translinear $\mathrm{eml}$ cell without scaffolding is not a cheaper cell but
a circuit whose defining relation has failed. The overhead therefore exceeds
unity for any realization in this family, and that much is structural. What
simulation supplies is the magnitude, and magnitude is where a redesign has
room; we make no claim that this cell is the best obtainable, and the
argument does not need one.

\emph{How much of the \num{3046} is architecturally recoverable?} That
freedom can be priced without a new circuit, and Supplementary Note~S5 does
it from Table~\ref{tab:blockbudget}: separating the supply
current every cell must replicate from the servo current that could in
principle be shared across cells gives
$F(N) = \num{3.84} + \num{11.24}/N$, which returns the shipped
$F = \num{15.1}$ at $N = 1$ and falls to \num{4.19} at $N = 32$. A shared
servo also needs one compensation capacitor rather than one per cell, so
per-cell $C$ falls toward the $\approx \SI{50}{\femto\farad}$ of parasitics.
Together at $N = 32$ those move $FC$ by \num{144}, which would put the
composability overhead near \num{21}$\times$ rather than \num{3046} and the
fabric-versus-digital comparison at \numrange{0.08}{2.8}$\times$, winning at the favorable end of the sensitivity box and still losing at the adverse one.

Three things that estimate does not license. It is arithmetic on the energy
law and not a circuit: sharing a servo across a column cannot tie the
collectors together, since each carries its own signal current, and needs a
per-cell cascode on a shared bias that nobody here has drawn. It holds
$N_\tau$ fixed, and a cascode that holds a collector less firmly than a
servoed node raises the precision demand that sets $N_\tau$, so the true
figure is worse than \num{144}. And it leaves the second requirement
untouched: the fabric would still resolve \num{2.3} bits where it needs about
\num{8}, and the matching that buys those bits costs area that returns as
capacitance in the same term. Sharing the scaffolding is the largest single
lever this analysis exposes, it is worth roughly two orders of magnitude, and
it is not by itself enough.

The precision argument has to be reconciled rather than refuted, and the
reconciliation is that Sarpeshkar priced a computing element. On that object
we confirm him: the bare core beats the modeled datapath deep inside the
regime he identified~\cite{sarpeshkar1998analog}. What the citation practice
adds, and what he did not claim, is that a machine assembled from such
elements inherits the advantage. It does not. The fabric operates nine to
eleven bits inside the regime where analog computation is supposed to win and loses there anyway. At \num{4.3} bits the shot-noise floor would permit
\SI{6.2e-17}{\joule} against the \SI{3.96e-11}{\joule} spent, so the
cell delivers the charge of a \num{13.9}-bit computation to produce four
bits. A low-precision regime is necessary for an analog fabric to be
competitive and nowhere near sufficient, because the term that decides the
outcome is not in the precision--energy trade at all. Supplementary Note~S7
gives the crossover arithmetic and a $kT/C$ hypothesis that fails in the same
direction; Supplementary Note~S6 disposes of two objections from the
log-domain literature, that a lower bias current recovers the difference and
that subthreshold parts dissipating
nanowatts~\cite{sarpeshkar2010ultralow,mandal2009logdomain} contradict these
figures. Neither does.

The fabric's second limit is upstream of any circuit, and it is a theorem
about a class rather than a quirk of one operator. Exponentials and
logarithms of real arguments are eventually monotone, and so is anything a
grammar assembles from finitely many of them, while sine never stops turning
around. The real closures generated by $\exp$, $\ln$, the algebraic
operations and $\arctan$ lie in a Hardy field, every element of which is
eventually monotone and therefore has finitely many zeros, and $\sin$ has
infinitely many zeros in every neighborhood of
$+\infty$~\cite{rosenlicht1983hardy,boshernitzan1982orders}. No real
primitive whose closure lies in that class represents trigonometry exactly,
at any depth. The equations Section~\ref{sec:res-grammar} reports as blocked
are therefore blocked for the whole class, and any repair must leave it by
introducing an oscillatory device, which brings its own domain, stability and
calibration costs rather than removing them. The scope of the theorem has to
be stated with it: it concerns germs at $+\infty$, while the benchmark
evaluates on bounded boxes, and on a compact domain the closure of $\{1,x\}$
under $\exp$, $\ln$ and the algebraic operations is dense in the continuous
functions. It does not say a trigonometric target cannot be
\emph{approximated} on a bounded box; it says the grammar does not
\emph{contain} it. The exclusion count belongs to the exact statement and the
$\sin(k \ln x)$ results to the empirical one, and neither is offered as
evidence for the other. Supplementary Note~S8 keeps them apart in full.

The complementary count is more damaging than the exclusion. Only about one
equation in nine of the corpus $\mathrm{eml}$ was proposed for contains
$\exp$, $\log$ or $\tanh$ at all, while nearly twice as many are excluded
outright. A primitive that is universal in the algebra is not thereby well
matched to the workload, and the gap between those two statements is not
recoverable by better circuit design. Consistent with that, every comparison
in which the fabric was given a resource-matched opponent was lost, including sensor linearization, the application an exponential--logarithmic machine should own, where a three- to four-parameter polynomial in $1/y$ against
$\log x$ is exact on both thermistor targets and the fabric needs
\numrange{92}{297} parameters to reach
\SIrange{0.009}{0.014}{\kelvin} ideal and \SIrange{0.033}{0.044}{\kelvin} on
the realized cell. Two comparisons in which it appears to do better are given
in Supplementary Note~S9, each with the half that removes the appearance.

Those comparisons all require the fabric to compute something particular at
every interior cell, and the obvious objection is that this is the wrong way
to use it. A machine running at two to six bits may be a poor approximator
and a serviceable random projection, which is the regime much of the physical
unconventional-computing literature
occupies~\cite{teuscher2023reservoir,markovic2020physics} and the one that relaxes the requirement we identify as binding, since a random
projection has no target value at any interior node.
Section~\ref{sec:res-elm} puts the fabric there. The first half of the answer
confirms the diagnosis from the behavioral side: the extracted
non-idealities cost the trained fabric \num{2.56}$\times$ and the frozen
fabric nothing at all. The same silicon is expensive when interior values
must mean something and free when they need only be repeatable. That
comparison contains no energy model, no digital baseline and no node-scaling
factor, so it is an independent route to the claim the overhead makes from
the charge side: what the scaffolding buys is the right of an interior value
to be believed.

The second half is that the regime does not save the fabric. Frozen on the
full silicon configuration it improves by an order of magnitude and still
loses to the resource-matched opponents, and against a textbook random-feature machine, a Gaussian projection through $\tanh$ at the same feature count and readout, it loses by roughly three to nine times. The
operator that is functionally complete for the elementary functions is a
worse random nonlinearity than the cheapest one available, which is the
sharpest form the workload-mismatch finding takes anywhere here. Nor can the
improvement be credited to the change of regime: a frozen fabric read out at
layer~0 alone, which is the readout the trained fabric uses, is level with
the trained fabric. The order of magnitude comes from reading every cell, and
an all-cell readout is one more output path per cell, the same per-cell peripheral the energy law charges for, though priced against
Table~\ref{tab:blockbudget} it is a cheap one, some \SI{3.5}{\percent} of
the cell, which buys the accuracy rather than cancelling it. The escape is
therefore real in energy terms and still bounded by the same architecture: it
is bought by adding a peripheral to every cell, which is the move the rest of
this argument says does not scale.

Changing the operator does not escape it either. A third, current-summed port
bought nothing, because the fabric already sums and its depth is limited by
mismatch rather than by expressiveness. Two better-conditioned operators
remove $\mathrm{eml}$'s domain hole and its unbounded $v$-port sensitivity at
no circuit cost, worsen dynamic range, leave trigonometry unreachable for the
reason just given, and have completeness \emph{conjectured} only, so each is a trade and not an upgrade, Odrzywo\l{}ek's theorem remaining the only
proved completeness result in the family. The Lambert-$W$ operating point
$u - W(e^{u+v})$ is better conditioned on every axis tested and provably
nonexpansive, and it does not remove the scaffolding, because a diode's
operating point still has to be held. Supplementary Note~S8 gives all three.
One limit we promised early turned out not to be the binding one: the port
asymmetry of \eqref{eq:ports} is real and shaped the design, but the per-hop
$v$-port attenuation we had treated as a second effect is the logarithmic
pedestal counted twice (Section~\ref{sec:res-nonideal}), and the depth this
fabric reaches is set by mismatch rather than by either port. The asymmetry
survives as a design constraint. It is not why the fabric fails.

The failure has a specific address, and it is not the operator. Per-cell
power would have to fall by roughly two orders of magnitude and per-cell
resolution rise from \num{2.3} bits to approximately \num{8}, together:
a cheap cell that resolves two bits computes nothing, and an accurate cell at
the present power buys accuracy digital already supplies for less. Whether
better hardware answers this depends on which requirement is asked. Precision
is in part a process property, since the \SI{20.7}{\percent} level sigma is a
simulated device mismatch and mismatch follows the area law of Pelgrom et
al.~\cite{pelgrom1989matching}, whose consequences for the
matching--area--power trade Kinget sets out~\cite{kinget2005mismatch}. Energy
is not a process property at all. Nor are the two independent: buying
precision through matching costs area as the inverse square of the sigma, so
the required improvement asks for roughly \num{50} times better matching and
some three orders of magnitude in device area, which returns as capacitance
in the term the energy law charges for. A better-matched process buys depth;
it does not buy energy. Both requirements point at the same circuitry, the servo scaffolding rather than the translinear core, and concretely at a composition mechanism that does not hold each cell's operating point
actively, a compensation strategy that does not put two picofarads in every
cell, and calibration shared across cells rather than stored per cell. A
design meeting all three would be a different architecture, and the win
available to a scaffolding-free successor is available only to a circuit that
is no longer a programmable fabric, since $F = 1$ means the collector
voltages float and the translinear law fails, and
$C \approx \SI{50}{\femto\farad}$ means an uncompensated variant that rings.

That requirement says nothing about what could meet it, and the search should
not be aimed at the exponential: intrinsic $\exp$ and $\ln$ are already
solved in this technology, so a material that merely implemented them
natively would tie this result rather than beat it. The overhead lives in $F$
and $C$, both of which exist to hold a node where it was put. The design
question is not which material computes $\mathrm{eml}$; it is which material
holds its own operating point. Nothing chemical, superconducting or
memristive was fabricated, simulated or extracted anywhere in this study.

Generalized, the overhead is the price of holding a value that is not an
attractor. Digital logic gets restoration free because its states are
attractors: a gate pulls its output to a rail and keeps pulling, so a
disturbance is undone by the same transistor that set it. An arbitrary value
in a continuum sits at no attractor, so something external must hold it, and
in a cascade that something is paid for again at every stage. Physics
supplies restoration, but only discretely. That a continuum of equilibria has
a singular Jacobian along itself, so a value held there drifts diffusively,
is the standard continuous-attractor result in computational
neuroscience~\cite{seung1996eyes,burak2012limits}, and that noise collapses
an analog machine to finitely many distinguishable states was proved for
analog neural networks by Maass and Sontag~\cite{maass1999analog}. Neither is
ours. What is new here is the price, measured on a netlist tied to a layout.
Restoration and continuous-valuedness are mutually exclusive for any
autonomous substrate, \emph{in the limit of indefinite holding}. We do not drop that qualifier, since computation needs a value held for a settling
window rather than forever, and a slow manifold with a small nonzero
restoring rate is not excluded by a linearization argument. Overdamped
Langevin simulation of five canonical landscapes agrees on both halves and
corrects a claim an earlier version of this argument made: a chemical steady
state does restore, but it holds \emph{its} value rather than an arbitrary
one, so it falls to the proposition rather than exempting itself from it. If
restoration is available only at isolated points, a substrate that holds a value holds one of finitely many, and a $b$-bit analog value needs $2^{b}$ wells, which is to say it has rebuilt digital restoration under another name. Supplementary Note~S10 gives the landscapes and a Kramers
estimate of what a barrier costs.

The obvious objection is a device that holds an analog value with no power at
all. Memristive memories are the strongest form of it, and they are an
instance of the proposition rather than an exception: retention is bought
with an energy barrier, so the surviving states are discrete and few and
programmed conductances drift between them, with open-loop analog storage
running to roughly four bits~\cite{zhao2020reliability}. What such a device
does establish, and a translinear node cannot match, is that holding can cost nothing once the value is in place. That is a different economy from the one priced here, where every cell pays for the duration of an evaluation, and the
reason in-memory architectures pay for conversion instead.

Nothing in what follows was designed, simulated or tested, and we offer it as
a reading of where the results point rather than as a result. Four routes are
open and the proposition prices them differently. The circuit-level route
attacks the magnitude and not the structure: $F$ falls if scaffolding is shared rather than replicated, and $C$ falls under a compensation strategy that does not put a Miller capacitor in every cell. Neither changes the sign of the comparison alone. Two alternatives we did not test belong here
because a reader will reach for them and we cannot price them: a subthreshold
CMOS realization would change the transconductance-to-current ratio, the loop
gain each node needs and therefore the compensation, and we have designed
neither it nor a self-biasing log-domain loop of the kind the companding and
translinear literature
develops~\cite{seevinck1991companding,gilbert1975translinear,frey1993logdomain},
in which the operating point is held by local device physics rather than by
an operational amplifier. That family is the most credible route to $F$ near unity because it removes the amplifier rather than sharing it, and
it is the variant we would most want priced against the energy law. What we would want reported for either, and for any analog
fabric offered as an alternative to digital, is $F$ and $C$: they are the
two variable terms of its energy law and neither is expensive to measure.

Composition is where we would look first, because the overhead is paid once per cell because every cell holds its value in space while its
neighbours hold theirs. A fabric that reuses one cell in time pays for
scaffolding once and converts a cost that does not amortize across cells into
one that amortizes across evaluations; a fabric that restores digitally
between analog stages buys the property matter supplies only discretely, at
one conversion per stage, which at two to six bits is not obviously the
larger cost. Neither is exotic and neither was tried here.

The material route is constrained rather than closed: what is ruled out is a
substrate that holds an arbitrary value handed to it from upstream, and what
is not is a substrate whose own attractor is the answer. The Lambert-$W$
operating point is the small version of this and relaxation networks the
large one. The cost moves rather than disappearing, since such a machine
computes by settling, and the attractor census found no physics-native
operator with more than one stable fixed point as a relaxing node. Whether a
coupled network escapes what a single node cannot is open, and it is the
question we would put to anyone offering a material as a computational
substrate.

A fourth route is already in the literature and dissolves the subtraction
rather than shrinking it. In floating-gate-enabled field-programmable analog
arrays the switches that route signals also store analog values and perform
vector--matrix multiplication, so the interconnect is a place where
computation happens rather than a levy on it~\cite{hasler2020fpaa}. An
overhead defined as cell energy over the energy of the part that computes is
not well posed on such a fabric, because the denominator is no longer
confined to a core. We take that as a limit on our metric rather than an
objection to it: the ratio is meaningful where a fabric divides into a
computing element and machinery that holds it, and a designer who can make
the holding machinery compute has left the regime it was built to describe.
Whether an $\mathrm{eml}$ primitive admits that treatment, given that its
ports must be held to within a fraction of a thermal voltage, we do not know.

The negative results concern this primitive and this realization. They
do not show that analog computation is uncompetitive in general, that
translinear circuits are a poor substrate, or that Odrzywo\l{}ek's
completeness result is of no consequence; the mathematics is
unaffected by what a \SI{130}{\nano\meter} implementation costs. They
do not show that a different $\mathrm{eml}$ circuit, in a bipolar
process with better matching or with a fundamentally different
composition mechanism, would fail. What generalizes is narrower: three
costs are incurred per cell and do not amortize: holding the primitive
at the precision its own nonlinearity demands, making its analog value
addressable and composable, and storing the per-instance calibration
that mismatch forces. The primitive itself amortizes to approximately
nothing. That reasoning transfers to any single-primitive analog
fabric whose cells must be individually held, addressed and
calibrated, and its practical consequence is that the question worth
asking of a candidate primitive is not how few devices evaluate it but
how tightly it must be held, whether its composition mechanism can be
passive, and whether its calibration can be shared. It does not
transfer to architectures that avoid those three costs by
construction, and we have not shown that no such architecture
exists. The overhead reported here is a measurement of one cell, not a
law. Nothing here bears on trainability either: every statement about
fabrics of more than four cells is about representability, and the
optimization results are gradient descent through a behavioral model
rather than in-situ learning on hardware. Three further limits belong
to the extension studies: the restoration proposition is proved for
autonomous dynamics and therefore covers storage, not the driven
composition this fabric performs; the better-conditioned operators and
the Lambert-$W$ point have completeness \emph{conjectured} only; and
the third-port and operator-comparison results are behavioral, so
their power figures are estimates from the cell's own current budget
rather than extracted quantities.

What is a universality theorem worth once someone has to build it?
Odrzywo\l{}ek proved that a single two-argument operator generates the
elementary functions, and we asked what a machine assembled from nothing
but that operator would cost and what it could do.

Changing how the fabric is used does not escape that conclusion either, and
the attempt is worth more than the result. Freezing the interior and fitting
only a linear readout removes the requirement that any interior value be
accurate, and the extracted non-idealities stop costing anything the moment
it is removed: \num{2.56}$\times$ trained, \num{0.95} frozen. The tax is the
price of a value that has to be believed. The machine that results is still
beaten by its resource-matched opponents, and beaten by a factor of three to
nine by a random projection through $\tanh$, so the operator is not a better random nonlinearity than the cheapest one, either.

Three attempts to escape that conclusion by changing the operator did not
escape it. A third additive port moved none of six targets to a shallower
depth, because the fabric already sums. Two better-conditioned operators close
$\mathrm{eml}$'s domain hole and bound its $v$-port sensitivity at no circuit
cost, but their completeness is conjectured and not proved, so the trade buys
conditioning at the price of the guarantee that motivated the operator in the
first place. The substrate route is narrowed by a further proposition: on any
autonomous substrate, restoration and continuous-valuedness are mutually
exclusive in the limit of indefinite holding, because a continuum of
equilibria has a singular Jacobian and hence exactly zero restoring rate.
Matter restores discretely, so a substrate that holds a value has rebuilt digital restoration under another name. That proposition is proved for
storage and is not ours~\cite{seung1996eyes,maass1999analog}; what is new is
its price. Whether it extends to driven composition, which is what this fabric
performs, is open.

What would have to change is specific. Per-cell power would have to fall
by roughly two orders of magnitude and per-cell resolution rise from
\num{2.3} bits to about \num{8}, together, and the circuitry to attack is
the \SI{74.5}{\percent} of supply current spent on servos rather than the
\SI{2.7}{\percent} of area spent on computing. We have not shown that no
architecture meets those conditions. We have shown that this one does
not, and why, in simulation, ahead of silicon. A universal primitive can
be nearly free and still yield an unusable fabric, because composability
rather than computation sets the cost. 

\section{Methods}
\label{sec:methods}

\subsection*{Evidentiary basis}
\label{meth:basis}

The design has
been submitted to a multi-project shuttle and is in fabrication, but no part
had returned at the time of writing and nothing reported here was obtained
from silicon.
Every quantity we report comes from one of three tiers of evidence, and we name
the tier at each use.

\emph{Tier 1, numerical and behavioral simulation.} Our primary evidence is a
differentiable model of the $\mathrm{eml}$ network, which we wrote in
PyTorch~\cite{paszke2019pytorch} and
evaluated in double precision, and into which we inject hardware-derived
non-idealities as explicit forward-model terms. All accuracy, capacity,
topology, baseline-comparison and dynamical-system results are of this kind.
Where we use the fabric as the nonlinear term of an ordinary differential
equation, \texttt{scipy} performs the integration and the fabric supplies only
the nonlinearity. We designed and simulated no analog integrator, and no
closed-loop result should be read as an analog dynamical system.

\emph{Tier 2, circuit-level simulation.} We obtain the parameters of the
non-ideality model, and every power, settling and energy quantity, by
\texttt{ngspice} simulation of netlists tied to the mask database by
layout-versus-schematic (LVS) equivalence or by parasitic extraction. Results
labeled \emph{post-layout} come from parasitic-extracted netlists; results
labeled \emph{schematic} come from the LVS reference netlists. We never mix the
two bases within a quantity.

\emph{Tier 3, physical layout verification.} We read geometric and connectivity quantities off the mask database or the placement record with a layout tool: areas, device censuses and rule-violation counts.

We normalize all currents to the unit current
$I_{\text{unit}}=\SI{0.5}{\micro\ampere}$, the on-chip bias reference, and write
one such unit as one \emph{unit}. We do not use the word \emph{measured} of
any quantity in this paper. Nothing here was measured; every number is
simulated, extracted or modeled in the senses just defined, and we say which.

\subsection*{Network model and training}
\label{meth:network}

\paragraph{The fabric as a differentiable model.} A fabric is a layered
arrangement of identical cells, each computing
$\mathrm{eml}(u,v)=\exp(u)-\ln(v)$ over the reals. An analog circuit has no
access to the complex plane, so the model is real-valued throughout and the
logarithmic input rails at a positive floor in the manner of a log amplifier.
Each of a cell's two ports takes a programmable affine combination of the
available signals,
\begin{equation}
u = \alpha_u + \textstyle\sum_{k}\beta_{u,k}\,x_k + \sum_{f}\gamma_{u,f}\,c_f ,
\qquad
v = \alpha_v + \textstyle\sum_{k}\beta_{v,k}\,x_k + \sum_{f}\gamma_{v,f}\,c_f ,
\end{equation}
where $x_k$ are the external input variables and $c_f$ the outputs of the cells
feeding this one. A full-precision digital readout stage $A\cdot\text{root}+B$
closes the model. In the single-variable case a cell therefore stores six numbers
($\alpha,\beta,\gamma$ per port) and the fabric stores two more, so a depth-$D$
binary tree of $2^D-1$ cells costs $6(2^D-1)+2$ stored parameters: 44 at depth 3,
92 at depth 4, 380 at depth 6 and 1532 at depth 8. This count is the resource
measure we use for every comparison in this paper.

\paragraph{Topologies.} We express connectivity as a per-layer gather table
rather than a hardwired index, and we used three arrangements. A \emph{tree} is
a full binary tree with no sharing, so cell count grows as $2^D$. A \emph{DAG} has
constant-width layers in which every cell may draw from every cell in the layer
below, so subexpressions are shared and cell count is linear in depth. A
\emph{mesh} is a DAG with a local connection window (window 3 = self plus two
neighbors), i.e.\ nearest-neighbor tile routing. We evaluate layers one at a
time, vectorized over cells.

\paragraph{Leaves and multivariate inputs.} The selectable leaves are the rails
$[1,x_1,\dots,x_n]$; the constant is a leaf on the same footing as a variable, so
calibration constants are learned rather than supplied. We used three
leaf-to-variable assignment schemes and report them separately. \emph{Dense}
gives every input path a coefficient on every rail, so the assignment is learned
and continuous. \emph{Soft} keeps one scalar gain per path and learns per-path
logits over the rails, read out by a softmax at temperature $\tau=1$ during
training and by an $\arg\max$ at evaluation, so the assignment is learned and
discrete. \emph{Search} fixes a hard random assignment per candidate and performs
a random search: 12 candidate fabrics, each trained for 250 iterations, the best
of which is retrained under the full training schedule.

\paragraph{Initialization.} Two initialization schemes appear in the record, and
we distinguish them everywhere, because the difference between them is larger
than most effects reported here. The \emph{legacy} scheme draws affine weights
from
$0.7\,\mathcal{N}(0,1)$ with the logarithmic port's constant biased by $+1.5$ to
keep its argument positive. The \emph{unity-gain} (identity) scheme instead
linearizes the cell about $u_0=-2$, $e_0=\exp(u_0)=0.135$ and $A=\exp(e_0)$, at
which the cell's constant term vanishes, and then asks each cell to be a
unity-gain average of its children by setting
$\gamma_{u,f}=w/e_0$, $\gamma_{v,f}=-wA$ with $w=1/(2F)$ over $F$ child
connections. We drive both paths, since routing only the logarithmic path
leaves a deep tree connected along a single root-to-leaf spine with almost every
cell dead. A stack of such cells is a bounded-gain pass-through rather than an
exponential tower, and that is what makes depth beyond 4 trainable; the
improvement is a factor of 33 at the median. We therefore qualify every
depth-versus-accuracy statement by the initialization used.

We then solve the readout in closed form by least squares against the target, so
training starts from the best linear readout of the fabric's initial features.
With a single scalar input the layer-0 features are correlated to about 0.999,
and plain least squares returned readout weights of order $5\times10^{4}$ and
fabric outputs of \SI{2349}{\kelvin} on a \SI{298}{\kelvin} target. We therefore
used a relative ridge term $\lambda=10^{-6}$ throughout.

\paragraph{Optimizer and replication.} We trained full-batch with
Adam~\cite{kingma2015adam}, learning
rate 0.02 decayed on a cosine schedule to 0.002, gradient-norm clipping at 1.0,
evaluation every 100 iterations, and retention of the best clean-evaluation state
rather than the final state. Finite weight resolution, where enabled, we trained
as quantization-aware training with a straight-through estimator. Iteration
counts were 3000 for the sensor, topology, ODE and contracting-system studies,
4000 for the depth sweep, 2500 for the multivariate study and 2000 for the
machine-learning study. Seed counts were 6 for the single-knob sensitivity sweep,
8 for the combined and PDK-calibrated chip runs, 4 for the ODE study and 3 for the
topology, machine-learning, contracting-system and multivariate studies. Training
sets were 768 points for the contracting-system study, 512 for the synthetic
machine-learning tasks and 256 for the scalar sensor tasks, each with a disjoint
in-range test set and, where applicable, a disjoint extrapolation set. Because
seed counts are small, we report both the median over seeds and the best seed,
and we state which. Where we compare two model families, we take best-of-family symmetrically, the best fabric of any admissible size against the best baseline of any admissible size, and we state the comparison criterion with the count, since pointwise fit error and closed-loop trajectory error do not rank
the models identically. All defaults reduce to ideal real-domain arithmetic,
verified to machine precision, and a fixed-seed regression file re-checks the
pre-existing scalar results bit-for-bit.

\subsection*{Non-ideality model}
\label{meth:nonideal}

Hardware-derived effects enter the network model as independently switchable
terms of an explicit forward model: an output and exponential-path saturation
ceiling, a logarithmic-input floor, per-stage additive noise redrawn every pass,
static per-cell gain and offset mismatch drawn once per chip from a mismatch
seed, finite affine-weight resolution and range, a global thermal-voltage drift
that scales the exponential argument by $1/(1+\delta)$ and the logarithmic output
by $(1+\delta)$, a small saturating rail leak, the logarithmic pedestal, a
per-hop attenuation of the $v$ port, and a finite device span with an optional
hinge penalty in decades.

\paragraph{The pedestal is the physical configuration.} Fixed bias pedestals in
the logarithmic path mean that the realized cell computes $\ln(v+s)$ rather than
$\ln v$. On the segmented-resistor basis, which is what the layout draws,
$s=\num{2.547}$ units, and the extracted cell law is
$i_o = \num{2.478} - \num{1.094}\ln(v+\num{2.547})$; on the lumped LVS-netlist
basis $s=\num{2.574}$ with slope \num{1.145}. The two bases differ by
\SI{1.05}{\percent} in $s$ and \SI{4.5}{\percent} in slope. All headline studies
use the segmented basis.

\paragraph{Reproducibility trap: the pedestal and the $v$-port attenuation are
the same physics.} The model also carries a per-hop $v$-port gain
$\lambda_v$, and its nominal value \num{0.2551} is \emph{derived} from the
pedestal as $\num{1.094}/(\num{1.741}+\num{2.547})=\num{0.2551}$. Enabling the
pedestal and the attenuation together therefore applies the same silicon effect
twice. A probe of the realized per-hop gain confirms this directly: ideal
\num{0.6284}; pedestal only \num{0.2551}, which reproduces the extracted value
exactly; attenuation only \num{0.6284}, i.e.\ no effect on this probe; both
together \num{0.0933}, which is $2.7\times$ too small. \textbf{The physical
configuration is pedestal-only}, and every headline study uses it. We retain the
double-counted combination in the code only as an explicitly opt-in
pessimistic bound, guarded by a runtime warning, and we label any result routed
through it a bound and not a silicon configuration.

\paragraph{The configurations, and which results use them.} Five hardware
configurations appear in this paper. They are not a single ladder, and two of
them must never be read as points on one.

\begin{table}[tb]
\caption{\textbf{The five hardware configurations, and which results use
each.} \emph{Pedestal} and \textsc{pdk} are the two that must not be
confused: the first carries the realized cell law and no device statistics,
the second carries device statistics and not the cell law. \emph{Silicon} is
their union. Parameter values are not repeated here. They differ between studies and are given by the study that uses them.}
\label{tab:configs}
\centering\small
\begin{tabular}{@{}p{0.13\linewidth}p{0.36\linewidth}p{0.43\linewidth}@{}}
\toprule
Configuration & What distinguishes it & Where it is used \\
\midrule
\emph{ideal} & nothing enabled; real-domain arithmetic
  & the grammar enumeration, the ideal-fabric baselines, and the ideal arm
    of the \num{16}-system and random-feature studies \\
\midrule[\cmidrulewidth]
\emph{pedestal} & the realized cell law $\ln(v+s)$
  & the non-ideality study and the pedestal arm of the \num{16}-system and
    random-feature studies; this is the \emph{physical} configuration \\
\midrule[\cmidrulewidth]
\emph{\textsc{pdk}} & device mismatch, read noise, output rail and finite
    weight resolution, \emph{without} the pedestal
  & the depth and topology scaling, and the results labelled
    \emph{\textsc{pdk}-realistic} \\
\midrule[\cmidrulewidth]
\emph{silicon} & the pedestal \emph{and} the \textsc{pdk} effects together,
    read noise off for comparability
  & the random-feature comparison of Section~\ref{sec:res-elm} and its
    trained counterpart \\
\midrule[\cmidrulewidth]
\emph{worst case} & the pedestal \emph{and} the $v$-port attenuation
    \emph{and} the span limit
  & nothing reported here. It double-counts the pedestal, so it is a
    pessimistic bound and never a silicon result \\
\bottomrule
\end{tabular}
\end{table}

Table~\ref{tab:configs} lists them. \emph{Silicon} is the only configuration
in which both the cell law and the device statistics are present at once, and
every quantitative claim below names the configuration it belongs to.

\paragraph{The device-span limit does not bind.} The exponentiating device spans
$\SIrange{5.80e-6}{1366}{}$ units, that is \num{8.372} decades. At that true
span the over-excursion is \num{0.00} decades on every workload examined: the
signal never exceeds the high rail. Under-excursion is non-zero but lies below
the low rail, where currents are negligible. Only an artificial two-decade stress
configuration produces a real over-excursion (\num{2.171} decades), and those
numbers are not silicon numbers. We consequently disable the span term in the
headline studies. This is a null result rather than an omission.

\paragraph{Two non-ideality ladders, never juxtaposed.} Ladder~A is the
device-effect ladder, ideal $\rightarrow$ pedestal ($s=\num{2.547}$), used by the
contracting-system, ODE, machine-learning and non-ideality studies. Ladder~B is
the PDK-calibrated chip: gain $\sigma=\num{0.03}$, offset $\sigma=\num{0.005}$,
per-stage noise $\num{3e-3}$, saturation 30 and 8-bit weights. Ladder~B contains
none of the pedestal, attenuation or span terms. The gain $\sigma=\num{0.03}$
supersedes an earlier $\num{0.02}$ assumption, and we label results on the older
basis as optimistic. Combined Ladder~A$+$B configurations exist and are marked
as such. Finally, the rail leak (\num{0.02}) exists to keep gradients alive
through the rails; it makes gradient training an optimistic proxy for in-situ
training on hardware, and we carry that caveat with every in-situ claim.

\subsection*{Corpus, benchmarks and physical design}
\label{meth:elsewhere}

Three parts of the method are given in full in the Supplementary Information
rather than here, because no claim in this paper depends on reading them. The
exhaustive enumeration of the grammar
$S \rightarrow 1 \mid x \mid \mathrm{eml}(S,S)$, the deduplication of trees by
the function computed rather than by syntactic shape, and the two
admissibility tests that define domain validity are in Supplementary
Note~S1. The benchmark targets and their sampling ranges are in Supplementary
Note~S2, including one caveat that belongs with every count drawn from the
physics corpus: the \num{100}-equation AI Feynman set was transcribed offline
and its variable sampling ranges are documented substitutes for the published
ones, verified for structure, dimensional homogeneity and sampler finiteness,
with independent physics checks on \num{54} of the \num{100} and the remaining
\num{46} resting on structure and dimensions alone. The trigonometric
exclusion count is robust to that substitution, since it depends on the
equations' operator sets and not on their ranges; the baseline accuracy
figures are not. The physical-design flow, the tooling, and the provenance of
every extracted netlist are in Supplementary Note~S4. That last is the
vehicle rather than the subject of this study: the design supplies
device-derived parameters and no argument here rests on it.

\subsection*{Baseline approximators}
\label{meth:baselines}

We fitted four conventional approximator families under one harness:
polynomials of degree $p$ in $d$ inputs, costing $\binom{p+d}{d}$ coefficients
and fitted by ridge regression with the penalty chosen by five-fold
cross-validation over 13 logarithmically spaced values; cubic B-splines with $m$
interior knots on a fixed uniform grid, costing $m+4$ coefficients; lookup tables
with $N$ entries per axis, costing $N^d$ stored values, fitted to the data by
least squares on a hat-function basis so that the table never inspects the true
function; and multilayer perceptrons, whose weights and biases are counted
exactly. We additionally offered the polynomial, spline and lookup-table fits
identity, $\log x$ and $1/y$ input and output transforms, so that a baseline
which happens to match the target's functional form is not artificially
handicapped. We score every method with the same metric in the task's own units
on the same three sets: the training points, a fresh in-range test set, and an
extrapolation set.

\paragraph{Resource matching.} We match all comparisons by \emph{stored parameter count}, what a chip or a microcontroller would have to hold, rather than by cell count, layer count or wall-clock time. In the
contracting-system study the matching is enforced in code: the fabric's own
parameter count is read back after construction and the baseline hidden width is
solved to match it, with the same seeds and the same 3000 Adam iterations. The
resulting pairs are an 8-cell fabric at 69--85 parameters against perceptrons of
70 and 85 parameters, and an 18-cell fabric at 163--199 parameters against
perceptrons of 163 and 201 parameters. We leave the polynomial baseline under-resourced at 10--110 parameters, and it is therefore
conservative. Where we assemble a comparison across files rather than running it
in a single harness, we say so.

\paragraph{Model selection is on a validation split, and this mattered.}
We select perceptron hyperparameters on a held-out validation slice
comprising \SI{25}{\percent} of the training split, over a grid of
$\{1,2\}$ hidden layers $\times$ $\{\tanh,\mathrm{relu}\}$ $\times$ three
learning rates $\{\num{3e-3},\num{1e-2},\num{3e-2}\}$, that is 12 configurations
with one fit each; the winner is then retrained on the full training split over
all seeds. This is the same protocol the polynomial baseline receives through
cross-validated ridge. \textbf{The first pass selected the grid winner on
training error, which picks the most overfitting member of the grid and
handicaps the baseline, and it produced a spurious fabric advantage that was
found and corrected.} The fingerprint was a training error of \num{0.0366}
against a test error of \num{0.8490} on the same fit. Correcting the selection
moved the perceptron's test NRMSE from \num{1.0866} to \num{0.7121} on
\texttt{diabetes}, which erases what had appeared as a $1.52\times$ fabric win;
from \num{0.8490} to \num{0.7521} on \texttt{wine\_proline}; and from
\num{1.2556} to \num{1.0254} on the largest single change in the synthetic
suite. We recomputed all 14 affected tasks. We did not regenerate the stale
train-selected files, so both sets of numbers exist in the archive; we report
only the validation-selected values. The fabric, by contrast, received no
hyperparameter grid at all: a fixed mesh of depth 4, width 8 and window 3, one
learning rate, one initialization, one configuration. The remaining asymmetry
therefore favors the \emph{baseline}, which was tuned on held-out validation
while the fabric was not tuned at all. The fabric's reported performance is on
that account a lower bound on what a search would give it, and every negative
result below is conservative in the same direction.

\subsection*{Energy model}
\label{meth:energy}

\begin{figure}[tp]
\centering
\resizebox{\textwidth}{!}{%
\begin{tikzpicture}[font=\sffamily,line width=0.7pt,>={Latex[length=1.8mm]}]
  \tikzset{
    blk/.style ={draw,fill=white,rounded corners=1pt,align=center,font=\scriptsize,
                 minimum height=8mm,inner sep=2pt},
    ped/.style ={draw=hlresc,fill=hlresc!8,rounded corners=1pt,align=center,
                 font=\scriptsize,text=hlresc!70!black,minimum height=6mm,inner sep=2pt},
    dot/.style ={circle,fill=black,inner sep=0pt,minimum size=1.5mm},
    wl/.style  ={font=\scriptsize,inner sep=1.5pt},
    grp/.style ={draw=black!45,dashed,rounded corners=2pt},
    pics/ota/.style n args={1}{code={
      \draw[fill=white,line width=0.7pt] (-0.42,-0.62)--(-0.42,0.62)--(0.72,0)--cycle;
      \node[font=\scriptsize] at (-0.06,0) {#1};
      \node[font=\tiny] at (-0.27,0.31) {$+$};
      \node[font=\tiny] at (-0.27,-0.31) {$-$};
      \coordinate (-inp) at (-0.42,0.31);
      \coordinate (-inn) at (-0.42,-0.31);
      \coordinate (-out) at (0.72,0);}},
  }

  \node[font=\scriptsize\bfseries,text=hlcell,anchor=west] at (-2.4,10.35)
       {exponential path};
  \node[wl,anchor=east] at (-0.45,8.0) {$I_u=u\,I_{\mathrm{unit}}$};
  \draw[->] (-0.4,8.0) -- (0.95,8.0);
  \node[dot] (nu) at (1.0,8.0) {};
  \node[wl,anchor=north] at (1.0,7.76) {$n_u$};
  \node[blk,text width=17mm] (QR) at (1.45,9.25)
       {reference transdiode $Q_R$\\[-1pt]\tiny $I_R$ from \texttt{pbias}};
  \pic (A1) at (2.95,9.55) {ota={$A_1$}};
  \coordinate (uy8)  at (0,8.0);
  \coordinate (uy10) at (0,10.10);
  \coordinate (uqw)  at ([xshift=-3.5mm]QR.west);
  \draw (1.0,8.0) -- (uqw |- uy8) -- (uqw |- uy10) -- (2.53,10.10) -- (A1-inp);
  \draw (QR.east) -- (A1-inn);
  \draw (A1-out) -- (4.4,9.55) -- (4.4,8.0);
  \node[dot] (vb2) at (4.4,8.0) {};
  \node[wl,anchor=south west] at (4.45,8.05) {$v_{b2}$};
  \draw (1.0,8.0) -- (1.75,8.0)
        decorate[decoration={zigzag,segment length=2.4mm,amplitude=0.9mm}]
        {-- (3.65,8.0)} -- (4.4,8.0);
  \node[wl,anchor=north,align=center] at (2.7,7.76) {$R_u=V_T/I_{\mathrm{unit}}$};
  \node[blk,text width=17mm] (Q2) at (6.4,8.0)
       {expo\-nen\-ti\-ating NPN $Q_2$\\[-1pt]\tiny base driven by $v_{b2}$};
  \draw[->] (4.4,8.0) -- (Q2.west);
  \node[blk,text width=14mm] (PM) at (9.0,8.0) {PMOS mirror\\[-1pt]\tiny $1{:}1$};
  \draw[->] (Q2.east) -- node[wl,above]{$I_R\,e^{u}$} (PM.west);
  \draw[->] (PM.east) -- (13.6,8.0) -- (13.6,5.60);
  \node[wl,anchor=south] at (11.4,8.10) {sources $I_{\exp}$};

  \node[font=\scriptsize\bfseries,text=hlmdac,anchor=west] at (-2.4,-0.75)
       {logarithmic path};
  \node[wl,anchor=east] at (-0.45,3.9) {$I_v=v\,I_{\mathrm{unit}}$};
  \draw[->] (-0.4,3.9) -- (0.95,3.9);
  \node[dot] (nv) at (1.0,3.9) {};
  \node[wl,anchor=south east] at (0.95,3.98) {$n_v$};
  \node[ped,text width=22mm] (PA) at (1.0,6.45)
       {\textbf{bias pedestal} XLPA\\[-1pt]\tiny 8 units, \SI{936}{\nano\ampere}, from \texttt{pbias}};
  \draw[->,hlresc] (PA.south) -- (1.0,3.95);

  \draw[grp] (1.25,2.35) rectangle (4.80,5.45);
  \node[font=\tiny,text=black!55,anchor=south west] at (2.25,5.52) {leg A (signal)};
  \node[blk,text width=15mm] (QA) at (2.15,3.05)
       {transdiode $Q_A$\\[-1pt]\tiny collector $n_v$, emitter $v_e$};
  \pic (A2) at (3.40,4.75) {ota={$A_2$}};
  \draw (1.0,3.9) -- (2.15,3.9) -- (QA.north);   %
  \node[dot] at (1.75,3.9) {};
  \draw (1.75,3.9) -- (1.75,5.06) -- (A2-inp);
  \node[wl,anchor=south east] at (2.92,4.53) {$V_{\mathrm{refb}}$};
  \draw[->] (2.45,4.44) -- (A2-inn);
  \draw (A2-out) -- (4.55,4.75);
  \node[dot] at (4.55,4.75) {};
  \draw[->] (4.55,4.75) -- (4.55,3.05) -- (QA.east);
  \node[wl,anchor=south west] at (4.62,4.85) {$v_{oa}$};

  \draw[grp] (1.25,-1.15) rectangle (4.80,2.15);
  \node[font=\tiny,text=black!55,anchor=north west] at (1.30,-1.24) {leg B (reference)};
  \node[blk,text width=15mm] (QB) at (2.15,-0.15)
       {reference transdiode $Q_B$\\[-1pt]\tiny collector $n_{vb}$, emitter $v_e$};
  \node[ped,text width=20mm] (PB) at (-0.10,0.95)
       {\textbf{bias pedestal} XLPB$+$XLVR\\[-1pt]\tiny 12 units, \SI{1394}{\nano\ampere}};
  \pic (A3) at (3.40,1.55) {ota={$A_3$}};
  \draw[->,hlresc] (PB.east) -- (1.70,0.95);
  \node[dot] at (1.75,0.95) {};
  \draw (1.75,0.95) -- (2.15,0.95) -- (QB.north);  %
  \draw (1.75,0.95) -- (1.75,1.86) -- (A3-inp);
  \node[wl,anchor=south east] at (2.92,1.33) {$V_{\mathrm{refb}}$};
  \draw[->] (2.45,1.24) -- (A3-inn);
  \draw (A3-out) -- (4.55,1.55);
  \node[dot] at (4.55,1.55) {};
  \draw[->] (4.55,1.55) -- (4.55,-0.15) -- (QB.east);
  \node[wl,anchor=south west] at (4.62,1.65) {$v_{ob}$};

  \draw (4.55,4.75) -- (5.90,4.75)
        decorate[decoration={zigzag,segment length=2.4mm,amplitude=0.9mm}]
        {-- (7.50,4.75)} -- (8.20,4.75);
  \node[wl,anchor=south] at (6.70,4.95) {$R_{\mathrm{LN}}=V_T/I_{\mathrm{unit}}$};
  \node[dot] (nsv) at (8.20,4.75) {};
  \node[wl,anchor=north west] at (8.30,4.63) {$n_{sv}$};
  \node[ped,text width=17mm] (LO2) at (8.20,6.45)
       {XLO$_2$ bias\\[-1pt]\tiny 16 units, \texttt{pbias}};
  \draw[->,hlresc] (LO2.south) -- (8.20,4.80);
  \pic (A4) at (8.95,3.05) {ota={$A_4$}};
  \draw (8.20,4.75) -- (8.20,3.36) -- (A4-inp);
  \draw (4.55,1.55) -- (7.95,1.55) -- (7.95,2.74) -- (A4-inn);
  \node[blk,text width=16mm] (MP) at (10.35,4.75) {native-$V_t$ NMOS pass $M_P$};
  \draw[->] (8.20,4.75) -- (MP.west);
  \draw[->] (A4-out) -- (10.35,3.05) -- node[wl,right]{gate $g_{\mathrm{pass}}$} (MP.south);
  \node[dot] (ndl) at (11.55,4.75) {};
  \node[wl,anchor=south] at (11.55,4.85) {$n_{dl}$};
  \node[blk,text width=14mm] (NM) at (12.75,4.75) {NMOS mirror\\[-1pt]\tiny $1{:}1$};
  \draw (MP.east) -- (NM.west);
  \draw[->] (13.6,5.55) -- (13.6,5.48) -- (12.75,5.48) -- (NM.north);
  \node[wl,anchor=north] at (12.75,4.05) {sinks $I_{\ln}$};

  \node[dot] (out) at (13.6,5.55) {};
  \node[ped,text width=17mm] (LO3) at (11.55,6.45)
       {XLO$_3$ output pedestal\\[-1pt]\tiny 16 units, \texttt{pbias}};
  \draw[->,hlresc] (LO3.east) -- (13.6,6.45) -- (13.6,5.60);
  \draw[->] (13.6,5.55) -- (14.85,5.55);
  \node[wl,anchor=west,align=left] at (14.90,5.55)
       {$I_{\mathrm{out}}$\\[-2pt]\tiny KCL at the output node};

  \node[draw=hlresc,fill=hlresc!7,rounded corners=2pt,font=\scriptsize,align=left,
        text width=58mm,inner sep=3.5pt,anchor=north west] at (-2.45,-1.70)
    {\textbf{\textcolor{hlresc!70!black}{Where the offset $s$ comes from.}}
     XLPA injects a fixed pedestal onto $n_v$, the \emph{same} node the signal current arrives on, and the reference leg is pedestalled $1.5\times$
     harder (12 units against 8). The transdiode pair therefore forms
     $\ln\!\big[(I_v+I_{\mathrm{ped}})/I_{\mathrm{ref}}\big]$, so the die computes
     $\ln(v+s)$ and not $\ln v$. Fitted on the netlist that LVS-matches the
     shipped layout: $2.478-1.094\ln(v+2.547)$, max residual
     $2.2\times10^{-4}$ units in simulation, segmented basis).};
  \node[draw=black!55,fill=black!3,rounded corners=2pt,font=\scriptsize,align=left,
        text width=72mm,inner sep=3.5pt,anchor=north west] at (4.45,-1.70)
    {\textbf{Why the servo amplifiers exist.}
     $A_1$ pins $n_u$ to $Q_R$'s $V_{BE}$, so the $u$ port is a virtual node: the
     input current sees a fixed voltage instead of an exponential one, and the
     signal appears only as $I_uR_u$ across $R_u$.
     $A_2$ and $A_3$ each hold their transdiode's \emph{collector} at
     $V_{\mathrm{refb}}$, so both legs take the logarithm at identical collector
     voltages and the Early effect cancels in the difference.
     $A_4$ forces $n_{sv}=v_{ob}$, so the whole $\Delta V_{BE}=v_{oa}-v_{ob}$, and nothing else, falls across $R_{\mathrm{LN}}$; the native-$V_t$ pass
     device carries that current with headroom a standard-$V_t$ device would not
     have.};
  \node[draw=black!60,rounded corners=2pt,fill=black!3,font=\scriptsize,
        align=center,inner sep=4pt,anchor=north west] at (12.55,-1.70)
       {output is the KCL difference\\ of the two path currents\\[3pt]
        $\displaystyle \frac{I_{\mathrm{out}}}{I_{\mathrm{unit}}}
         \;=\; a\,e^{\,u} \;-\; k\ln(v+s)$};
\end{tikzpicture}}
\caption{\textbf{The $\mathrm{eml}$ cell, and where its cost sits.}
\emph{Circuit.} The two current-mode paths of one cell and the four servo
amplifiers that hold their operating points. In the exponential path a
reference transdiode $Q_R$ sets $V_{BE,\mathrm{ref}}$ and transimpedance servo
$A_1$ holds the $u$ port there, so the input current appears across $R_u$ and
drives the exponentiating NPN $Q_2$. In the logarithmic path two base-driven
transdiodes $Q_A,Q_B$, each held at $V_{\mathrm{refb}}$ by a servo ($A_2,A_3$),
present $\Delta V_{BE}$ across $R_{\mathrm{LN}}$, which servo $A_4$ converts
back to a current. The output is the difference of the two path currents, taken
by Kirchhoff's current law at one node, so the subtraction costs no devices.
\emph{Pedestals.} The orange blocks are fixed bias pedestals. Because XLPA
lands on the signal node $n_v$ while the reference leg is pedestalled
$1.5\times$ harder, the argument of the logarithm is $v+s$ rather than $v$. It
is a property of the circuit, not a fitting convenience, and it is the one
device effect carried into every network result reported here.
\emph{Scaffolding.} Only the four transdiodes and four resistors evaluate
$\mathrm{eml}$; everything else exists to hold those eight devices at their
operating points, and is the origin of the energy ratios in Results.
\emph{Basis.} Topology and unit ratios are transcribed from the netlist that
LVS-matches the submitted layout; the pedestal currents and the transfer
$i_o = 2.478 - 1.094\ln(v+2.547)$ come from DC sweeps of that netlist in
circuit simulation. No device was fabricated.}
\label{fig:cell}
\end{figure}
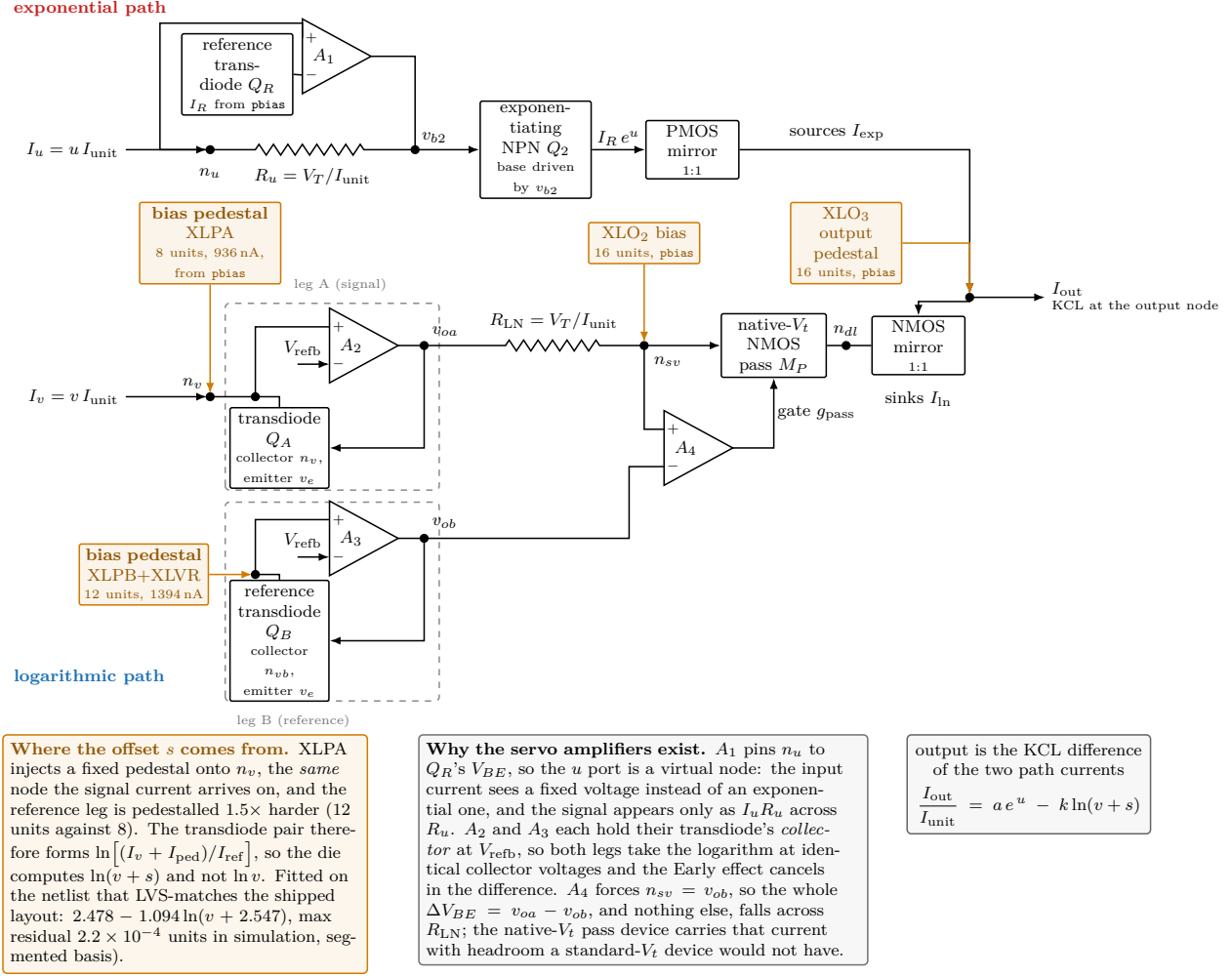

\paragraph{Derivation.} The cell the law describes is shown in
Figure~\ref{fig:cell}. We model the energy of one cell operation as
\begin{equation}
E_{\text{op}} = F \, N_\tau \, C \, V_{dd} \, n V_T ,
\qquad F = I_{vdd}/I_{\text{branch}},
\label{eq:eop}
\end{equation}
where $F$ is the ratio of total supply current to the current in the branch that
does the settling, $N_\tau$ the number of time constants to the required
accuracy, $C$ the dominant node capacitance and $nV_T$ the thermal voltage times
the subthreshold slope factor. The bias current cancels exactly: it appears once
in the numerator, through power $\propto I$, and once in the denominator, through
settling time $\propto 1/I$ since $g_m = I/(nV_T)$. The resulting form is
$C V_{dd} (n V_T)$ and not $CV^2$, because analog settling is exponential and the
second voltage factor is the thermal voltage times the number of e-folds
demanded.

\paragraph{Inputs and derived quantities.} From circuit simulation of one cell in
die context at \texttt{tt} and \SI{27}{\celsius}: $V_{dd}=\SI{3.3}{\volt}$,
$I_{vdd}=\SI{29.8686}{\micro\ampere}$, exponential-servo tail current
$\SI{3.96}{\micro\ampere}$, Miller compensation capacitance $C=\SI{2.00}{\pico\farad}$,
post-layout \SI{1}{\percent} settling time \SI{402}{\nano\second} and time
constant $\tau=\SI{45.9}{\nano\second}$. From these,
$F=\num{29.8686}/(\num{3.96}/2)=\num{15.1}$, the factor of two being the
differential input pair of the exponential servo, across which the tail current
divides so that the branch performing the settling carries half of it; and
$N_\tau=402/45.9=\num{8.8}$; neither is fitted. We assume two constants rather
than extracting them: $n=\num{1.3}$, the typical subthreshold slope factor of the
sky130 \SI{5}{\volt} NFET, which enters $E_{\text{op}}$ linearly, and
$V_T=\SI{25.85}{\milli\volt}$ at \SI{300}{\kelvin}. Used as a forward prediction
with nothing fitted, equation~\eqref{eq:eop} gives $\tau=\SI{33.9}{\nano\second}$
against \SI{45.9}{\nano\second} in simulation and $E=\SI{29.3}{\pico\joule}$
against \SI{39.6}{\pico\joule} extracted, both ratios \num{0.74}: the derivation
is confirmed to \SI{26}{\percent}.

\paragraph{Bias sweep.} We tested the claimed bias-invariance rather than
assuming it, with a dedicated \texttt{ngspice} campaign at sky130A \texttt{tt} and
\SI{27}{\celsius} with \texttt{method=gear}, \texttt{maxord=2} and a
\SI{2}{\nano\second} grid, holding the same die context and the same normalized
operating point at every bias. We ran four bias points, \num{0.125}, \num{0.250}, \num{0.500} (the shipped value) and \SI{1.000}{\micro\ampere}, and recorded supply current, power, \SI{1}{\percent} settling time and the
extracted time constant at each. We then computed piecewise log-log exponents in
$E \sim I^{p}$ between adjacent points. This table is a simulation result,
transcribed as a literal into the energy analysis. It is on the \emph{schematic}
basis; the headline energy and settling figures are post-layout, and the ratio
between the two settling values, $402/342.9=\num{1.17}$, is consistent with the
independently obtained \SI{+19}{\percent} parasitic inflation factor. We report
the two bases separately.

\paragraph{Digital comparator.} The digital side is a model, not a simulation of
a digital design. Horowitz reports per-operation energies in his ISSCC 2014
table~\cite{horowitz2014computing} at
\SI{45}{\nano\meter} and \SI{0.9}{\volt} (8-bit add \SI{0.03}{\pico\joule},
8-bit multiply \SI{0.20}{\pico\joule}, 32-bit multiply \SI{3.10}{\pico\joule}),
and we multiply those by a control, clock and operand-delivery overhead factor
$\mathrm{CTRL}$ and by a process-node factor
$K_{\text{node}}=(130/45)(1.8/0.9)^2=\num{11.56}$, with SRAM at
$1.41\,\mathrm{kB}^{0.617}$ \si{\pico\joule} and data converters at a figure of
merit of \SI{34}{\femto\joule} per conversion-step.

\paragraph{Sensitivity to the externally set constants.} Three constants are set
outside the simulation, and we recomputed each headline across the plausible
range of all three. (i) \emph{Accuracy in bits}: we compared six sourceable anchors. The first is \num{2.28} bits from a \num{1.288}-unit swing against a
\num{0.266}-unit level standard deviation, \num{3.00} bits at the root of a
63-cell depth-6 tree, \num{3.86} bits from the PDK-calibrated in-situ median
error of \SI{5.5}{\kelvin} over an \SI{80}{\kelvin} span, \num{4.30} bits,
\num{5.73} bits from a \SI{1.88}{\percent} untrimmable logarithmic-slope
mismatch, and \num{6.70} bits from the best of the eight seeds of the
PDK-calibrated chip run. These six do not measure one quantity. Two are
device-level ratios of swing to noise, three are task errors over a
temperature span, and one is an untrimmable ceiling; we sweep them together
because each has been used somewhere as the fabric's precision, not because
they are interchangeable, and only the device-level pair bounds what a cell
can resolve. The band \numrange{2.28}{5.73} bits quoted throughout is that
pair together with the mismatch ceiling. We traced the \num{4.30} figure
used in earlier scripts, during this work, to the single best of the nine ideal-hardware runs, three seeds by three variable-selection modes, on one nine-variable target, and it is therefore not a hardware
precision; the median-based \num{3.86} is the recommended anchor and we carry
the range through. Accuracy reaches a headline only through a width-matched
digital baseline. (ii) \emph{Digital control overhead}: $\mathrm{CTRL}\in[2,10]$
with core range $[2,5]$, the lower bound set by a hardwired piecewise-linear
block with no register file, in the sense of Balfour et al.'s energy accounting
for embedded datapaths~\cite{balfour2008energy}, and the upper bound by Hameed et
al.'s finding that functional-unit energy is under \SI{10}{\percent} of the
total in a customized datapath~\cite{hameed2010understanding}. The value 3.0
used throughout sits at the low end and is thus the least favorable admissible
choice for the fabric. (iii) \emph{Process-node scaling}:
$K_{\text{node}}\in[5.1,11.6]$, the lower end corresponding to a
\SI{1.2}{\volt} contemporaneous \SI{130}{\nano\meter} core rail and the upper to
the \SI{1.8}{\volt} sky130 standard-cell rail, with an empirical per-generation
cross-check landing between them. The value \num{11.56} used throughout is the
top of the range and is thus the most favorable admissible choice for the
fabric. Ratios involving the digital baseline move with these constants; the
ratio of the assembled fabric to its own bare translinear core does not, because
all three cancel from it.

\paragraph{Per-cell power and scaling laws.} Two per-cell power figures are on
record and we use both, each in its own context: \SI{98.566}{\micro\watt} for
one isolated cell in die context at \texttt{tt}/\SI{27}{\celsius}, and
\SI{96.95}{\micro\watt} as the per-cell coefficient of the two-cell chain law
$P(N)=\num{96.95}N+\num{85.64}(N-1)+\num{2.09}$ \si{\micro\watt}, which is the
value that reproduces the simulated two-cell total of \SI{281.622}{\micro\watt}.
Both exclude weight-DAC switching power, which we did not simulate at non-zero
codes, and both are therefore lower bounds. We model latency as
\SI{166}{\nano\second} for the first stage plus \SI{65}{\nano\second} per
additional stage, the latter established from only two points and flagged as
weak, inflated by \num{1.19} for parasitics. Area is
$A(N)=\num{24972}+\num{23550}N$ \si{\micro\meter\squared}, which we built by
taking each block's bounding box once and counting instances rather than by
extrapolating an aggregate.

\subsection*{Statistics}
\label{meth:stats}

Every fabric result is best-of-$n$ seeds selected on training error and
evaluated on held-out data, with $n = 3$ throughout except the
random-feature study of Section~\ref{sec:res-elm}, which used $n = 5$;
baselines receive the same seed count and the same optimizer budget as the
fabric they are matched against. Quantities reported over the \num{16}-system
suite are medians, and we name the criterion, pointwise fit or closed-loop trajectory, wherever a win count is given, because the two
disagree. Supplementary Note~S3 gives the seed policy, the best-of rule and
the regression guard in full.

\subsection*{Use of artificial intelligence}
\label{meth:llm}

An AI coding assistant (Claude, Anthropic) was used extensively and at every
stage of this work: to write the layout generators, the verification and
extraction scripts and the behavioral simulation code; to construct, run and
summarize the simulation campaigns, including the random-feature study of
Section~\ref{sec:res-elm}; and in preparing this manuscript.

No figure content was generated by AI. The three plotted figures are produced
from simulation output by \texttt{gen\_figures\_v5.py} in the released
repository, the circuit schematic is drawn from the netlist topology, and this
paper contains no photorealistic or synthesized imagery.

The choice of experiments, the interpretation of their results and every
claim made from them are the author's, as is responsibility for the whole of
this paper. No AI system is an author of this work, and accountability for
its content is not delegated.

\section*{Data availability}

The design files that support the findings of this study are openly
available in the project's Tiny Tapeout submission repository at
\url{https://github.com/christofteuscher/tt-eml-fabric}, which contains
the GDSII layout, the abstract views, the netlist and layout generators,
the implementation flow configuration and the accompanying documentation
for the \texttt{tt\_um\_teuscher\_eml\_fabric} macro. The same
repository also holds the simulation code and benchmark harness that
produced the network-scale results (\texttt{sim/}), and the
\texttt{ngspice} decks, netlists and post-layout extractions that
produced the circuit-level results (\texttt{spice/}), together with the
derived datasets generated and analyzed during this study. The design is in
fabrication on a multi-project shuttle; no part had returned at the time of
writing, so this study reports no silicon measurements and there is no
measurement dataset to deposit.

\section*{Code availability}
\label{meth:code}

The design repository is public at
\url{https://github.com/christofteuscher/tt-eml-fabric}. It contains the analog
layout as Python generators that emit the GDSII stream and the matching SPICE
LVS references from one source, together with the verification wrappers, the
flattening and extraction scripts, the LVS cross-reference reader and the shuttle
submission metadata and license. Re-running the generators reproduces the
submitted macro byte-for-byte apart from GDSII date fields, verified by a flat
per-layer XOR that is empty on all 39 layers.

The same repository holds the software and circuit-level artifacts behind the
results reported here. \texttt{sim/} contains the differentiable network model,
the expression-corpus enumerator, the benchmark transcription and its
verification script, the baseline harness, the energy analysis that produces the
composability overhead and the sensitivity box, the derived datasets and the
fixed-seed regression tests. \texttt{spice/} contains the \texttt{ngspice}
decks, the cell and chain netlists, and the post-layout extractions carrying
parasitic resistance and coupling capacitance. The decks reference the process
design kit through the \texttt{PDK\_ROOT} environment variable and reference one
another by paths relative to the deck being run, so they execute unmodified on
any machine with an \texttt{sky130A} installation.

\section*{Supplementary information}

This document contains twelve Supplementary Notes, S1--S12, beginning
after the references. They carry the corroborating detail, provenance and
per-item tables for claims argued in the main text; no argument in the
Article depends on reading them, and nothing in them is a measurement of a
physical device.

\section*{Acknowledgements}

The author thanks Andrzej Odrzywo\l{}ek for the result that motivates
this work, the Tiny Tapeout team and Efabless for the open
multi-project-wafer infrastructure, and SkyWater Technology and Google
for releasing the SKY130 process design kit. All layout, extraction,
verification and simulation used open tools. This work was supported by the
National Science Foundation under grant no.\ 2318139.

\section*{Author contributions}

C.T. conceived the study, designed and verified the circuit, performed
all simulations and analyses, and wrote the manuscript.

\section*{Competing interests}

The author declares no competing interests.

{\sloppy\emergencystretch=4em
\printbibliography}

\clearpage

\setcounter{section}{0}
\setcounter{figure}{0}
\setcounter{table}{0}
\setcounter{equation}{0}
\renewcommand{\thesection}{S\arabic{section}}
\renewcommand{\thefigure}{S\arabic{figure}}
\renewcommand{\thetable}{S\arabic{table}}
\renewcommand{\theequation}{S\arabic{equation}}
\captionsetup[figure]{name=Supplementary Figure}
\captionsetup[table]{name=Supplementary Table}

\begin{center}
{\LARGE\bfseries Supplementary Information}\\[0.6em]
{\large Composability rather than computation sets the cost of an analog
\textsc{eml} hardware fabric}
\end{center}
\vspace{1.2em}

\noindent
This section supports the Article above. It carries the
corroborating detail, provenance and per-item tables for claims argued in
the main text; nothing here is required to follow the argument, and nothing
here is a measurement of a physical device. Section numbers are cited from
the main text as \emph{Supplementary Note}~S1--S12.

\note{Expression-corpus enumeration}{si:s01}

How the grammar was enumerated, how two trees were judged to compute the same function, and the two admissibility tests that define domain validity. The main text quotes the resulting class counts; this note is how they were obtained.

To bound what an $N$-cell fabric can express we enumerated the grammar
$S \rightarrow 1 \mid x \mid \mathrm{eml}(S,S)$ exhaustively. We deduplicated
trees \emph{by the function computed} rather than by syntactic shape: two
trees are the same corpus entry if and only if their values agree to 9
significant digits at all 9 points of the grid
$x \in \{0.35, 0.5, 0.75, 1.0, 1.5, 2.0, 3.0, 4.5, 6.5\}$. We implemented the
enumeration as a dynamic program over equivalence classes rather than over
syntactic trees, which makes the class count exact. Depth $\leq 4$ is exhaustive
and yields \num{44559} distinct function classes. Depth 5 has \num{4.37e12}
syntactic trees and cannot be enumerated; a deliberately biased targeted round
over a \num{6264}-class subset supplies a lower bound only, and we use it nowhere
as a distribution statistic. Depth is the number of $\mathrm{eml}$ nodes on the
longest root-to-leaf path and equals the physical cell count.

Two admissibility tests then defined domain validity. A composition is
\emph{domain-dead} if any $\mathrm{eml}$ node receives a non-positive second
argument at any grid point, since the logarithm is then undefined. A composition
is \emph{out of range} if any internal node's current leaves the span of the
exponentiating device, taken from a Gummel sweep of the stock
\texttt{npn\_05v5} as \SIrange{2.90e-12}{6.83e-4}{\ampere}, that is
\SIrange{5.80e-6}{1366}{} units at $I_{\text{unit}}$. We evaluated surviving
classes through the extracted cell law with the affine port weights free, and
took residuals after an affine trim, so the reported quantity is a shape
error and not an offset error. The machine-readable corpus that the network
simulation consumes contains \num{6681} entries. This analysis is DC only, at
\texttt{tt} and \SI{27}{\celsius}, and we ran no Monte Carlo at $N>1$.

\note{Benchmark construction}{si:s02}

The targets, their sampling ranges, and the transcription of the AI Feynman set together with the verification it did and did not receive.

The multivariate benchmark is the AI Feynman equation set~\cite{udrescu2020aifeynman}.
We \emph{transcribed the equations offline} from the published physics rather
than downloading them, and then verified them by three checks independent of the
transcription itself. The structural check requires that every symbol appearing
in an expression is declared and every declared variable is used, which catches
dropped and renamed terms. The dimensional check propagates SI dimensions through
the expression tree and requires homogeneity, agreement with the declared
dimension of the left-hand side, and dimensionless arguments to every
transcendental function; we declare the dimensions from what each symbol
\emph{means}, not from the formula, so a wrong power, a missing $c^2$, a swapped
numerator and denominator or a lost permittivity all fail here. The physics check
consists of hand-written limiting cases, cross-equation identities and known SI
numerical values. A sampler smoke test additionally requires finite, non-constant
output over the declared ranges. Of 100 equations, 100 were kept and 0 dropped;
41 physics checks were run with 0 failures, covering 54 equations, so
\emph{the remaining 46 rest on structure and dimensions only}. We tagged
equations by the operations they require: 82 are representable by exponential and
logarithmic composition and 18 are trigonometry-blocked. The dimensionality
histogram is 1 equation in 1 variable, 15 in 2, 36 in 3, 27 in 4, 13 in 5, 6 in
6, 1 in 8 and 1 in 9. The baseline sweep runs a fixed 12-equation cross-section
rather than all 100, and we label it as such wherever we use it.

We state two limitations plainly, because they bound what the benchmark can
support. First, \textbf{the per-variable sampling ranges are documented
substitutes, not the published ones}: we could not obtain the AI Feynman
per-variable ranges offline, so each variable carries a range with an explicit
provenance code recording how we chose it: a physically meaningful interval, an interval keeping a denominator or difference bounded away from zero, or an
interval narrowed from a generic choice because the generic choice makes the
output span more than 20 decades, which is a sampling artifact rather than
physics. Sampling is independent uniform over the resulting box. Absolute error
levels are therefore not comparable with published AI Feynman results; only
within-benchmark comparisons between models on identical samples are.
Second, \textbf{dimensional analysis cannot detect a wrong dimensionless
prefactor}. A transcription error of that kind is invisible to the structural and
dimensional checks and is caught only by the hand-written physics checks, which
cover 54 of the 100 equations.

\note{Statistics, seeds and reproducibility}{si:s03}

How many seeds each study used, how the best-of rule was applied, which quantities are medians and which are single runs, and what is pinned by the regression guard.

Every distribution reported in this paper carries its sample size, a measure of
center and a measure of variability, as follows.

\emph{Network-simulation studies.} The unit of replication is the random seed,
which sets both the parameter initialization and, where enabled, the per-chip
mismatch draw. Seed counts are $n=6$ for the single-knob sensitivity sweep, $n=8$
for the combined and PDK-calibrated chip runs, $n=4$ for the ODE study and $n=3$
for the topology, machine-learning, contracting-system and multivariate studies.
At these sample sizes we report the \emph{median} over seeds as the measure of
center and the \emph{best} and, where informative, the \emph{worst} seed as the
measure of spread; we quote no standard deviation over seeds, because at $n=3$
to $n=8$ it would not be resolved. We give both the median and the best seed
for every headline, since they can differ by an order of magnitude and the
distinction changes the interpretation.

\emph{Circuit-simulation statistics.} All device-level statistical results come
from a single Monte Carlo campaign on the die-matched lumped-basis netlist
comprising 441 \texttt{ngspice} runs: one nominal \texttt{tt} reference, 200
samples at \texttt{tt\_mm} and 60 at each of \texttt{ss\_mm}, \texttt{ff\_mm},
\texttt{sf\_mm} and \texttt{fs\_mm}, for a pooled mismatch sample of $n=440$. All
441 converged and all were parsed. Pooled results use the arithmetic mean as the
measure of center and the sample standard deviation $\sigma$ as the measure of
variability, quoted both in absolute units and as a coefficient of variation,
with sample minima and maxima where the tail matters. A separate paired run at
two operating points reused the same 440 instances, so the two points are paired
rather than independent. The campaign varied \emph{device mismatch only}, through
the PDK's per-instance mismatch switch. It did not vary the global process corner
as a continuous distribution. The sky130A global-process Monte Carlo facility bins at point geometries only and none of this cell's geometries falls in a bin, so four discrete skew points represent process and we quote no process
standard deviation. It did not vary temperature; every Monte Carlo deck
ran at \SI{27}{\celsius}. Whether mismatch spread and thermal drift add or
partially cancel is therefore not established by this work, and we never stack
the corner campaign (nominal devices, mismatch off) and the mismatch campaign.

\emph{Resolution limits.} At $n=60$ per skew corner the $1\sigma$ uncertainty on
an estimated $\sigma$ is approximately \SI{9}{\percent}, so per-corner standard
deviations should be read as $\pm\SI{10}{\percent}$ and small corner-to-corner
differences in $\sigma$ are not resolved. A pooled sample of 440 resolves the
distribution to roughly $\pm 2.9\sigma$; any statement about a band edge further
out is a Gaussian extrapolation from the fitted $\sigma$ and we label it as such,
because nothing in this work establishes a $3\sigma$ or $6\sigma$ yield.

\textbf{We performed no hypothesis tests and we report no $P$ values.} All
comparisons in this paper are descriptive comparisons of simulated distributions,
fitted coefficients and resource-matched error metrics. Accordingly we use the
word ``significant'' only in its statistical sense, which arises nowhere in our
results; we describe magnitudes as substantial or considerable.

\note{Circuit design, tooling and netlist provenance}{si:s04}

The physical-design flow and where each netlist came from. This is the vehicle rather than the subject: no argument in the main text rests on the design, which supplies device-derived parameters and nothing else.

\subsection*{Circuit design, PDK and tooling}

The cell that supplies the non-ideality parameters evaluates
$\mathrm{eml}(u,v)$ in the current domain, so Kirchhoff's current law performs
the subtraction at a single summing node and it costs no devices. In the
exponential path a diode-connected reference NPN
(\texttt{sky130\_fd\_pr\_\_npn\_05v5}) sets a base voltage at the reference
current and a two-stage Miller-compensated amplifier servos the input node to
it, so the input current appears as a $\Delta V_{BE}$ at a second, nominally
identical NPN. The exponent scale is the single product
$\nu = R_u I_{\text{unit}} = \SI{26.9}{\milli\volt}$, so one unit of input
current is one thermal-voltage e-fold by construction. In the logarithmic path
two base-driven transdiodes, a signal leg and a reference leg, each carry a
servo amplifier holding its collector at a fixed potential to suppress the
Early effect; their base voltages are differenced across a second resistor and
a third servo converts that difference into a current. Four servo amplifiers
are therefore required, one per servoed node, and this is the dominant cost of
the cell: every servoed node is an argument of an exponential, so at the
extracted $nV_T=\SI{27.64}{\milli\volt}$ per e-fold a servo error of
\SI{0.28}{\milli\volt} is already a \SI{1}{\percent} output error.
Figure~4 of the main text gives both paths, the four servos and the bias pedestals
that make the logarithm's argument $v+s$ rather than $v$.

\paragraph{The tile.} The vehicle is a Tiny Tapeout SKY26c $2\times2$ analog
tile on sky130A occupying \SI{72072}{\micro\meter\squared}: two
$\mathrm{eml}$ cells of \SI{93.70}{\micro\meter} by
\SI{116.05}{\micro\meter} each, a radix-4 signed current MDAC on every port,
a proportional-to-absolute-temperature bias core and a 29-bit scan chain. The
two cells occupy a third of the tile and the infrastructure that makes them
usable occupies the rest.

The design targets the open SkyWater SKY130 process, PDK variant sky130A, in
its \SI{130}{\nano\meter} node~\cite{skywater2022pdk}. The vehicle is a Tiny
Tapeout SKY26c $2\times2$ analog tile~\cite{venn2024tinytapeout} of
\SI{72072}{\micro\meter\squared} carrying two $\mathrm{eml}$ cells, a radix-4
signed current-steering multiplying DAC on every port, a
proportional-to-absolute-temperature bias core and a 29-bit scan chain. We
simulated with \texttt{ngspice} release \num{46}, built with the KLU direct
linear solver~\cite{ngspice}, and did layout, geometric verification and
parasitic extraction in KLayout and Magic.
Device models were taken from
\texttt{\$PDK\_ROOT/sky130A/libs.tech/ngspice/sky130.lib.spice}. We identify
the process design kit by variant and not by revision, because none of our
run artifacts records one: the extractor writes only the technology name into
a netlist, and the only machine-written kit path in the run record belongs to
the digital wrapper flow rather than to the analog extraction or to the
circuit simulations. A reader reproducing this work should expect small
differences with kit revision and should record the revision they use, which
is advice we are in no position to have followed. All analog layout
was produced by programmatic generation rather than by interactive capture:
Python generators emit placement, routing and the GDSII stream directly, and
the same generators emit the SPICE reference netlists used for LVS, so layout,
LVS reference and simulation netlist derive from one source of truth.

\paragraph{Netlists, resistor basis and analyses.} We used two die-matched
schematic netlists and two parasitic-extracted ones. The segmented-resistor
netlist carries the scaling resistors as series strings and is the basis for
the transfer-function coefficients and the corner campaign; the lumped netlist
carries them as single bodies and is the basis for the power, settling and
Monte Carlo results. The two differ by \SI{4.31}{\percent} in resistance, so
every coefficient carries its basis and we never combine coefficients across
them; relative quantities such as a fractional standard deviation are
basis-independent. We resolved supply current by inserting a zero-volt series
source in every rail-connected device terminal, so each leg's current is
metered rather than inferred, and Kirchhoff's law closes on the instrumented
cell to \SI{0.045}{\percent}. Transient analyses started from the converged DC
operating point under a step-settling protocol and a kick-response stability
protocol. We validated every run by its returned point count rather than by
the simulator exit code, because \texttt{ngspice} can exit successfully on a
truncated run. The corner campaign spans 5 process corners crossed with 5
temperatures under 3 analyses, 75 runs, all complete. On the assembled chain 8
of 9 conditions are usable; \texttt{ss}/\SI{-40}{\celsius} failed to integrate
under both trapezoidal and Gear methods and is reported as unavailable rather
than as a pass.

\paragraph{Fitting.} We obtained transfer-function coefficients by unweighted
nonlinear least squares on the simulated DC sweep, fitting the logarithmic
path to $i_o = a_{\ln} - k\ln(v+s)$ with the pedestal offset $s$ \emph{free};
fitting with $s$ pinned, or fitting to $\ln v$, changes the recovered $k$ and
makes a sound cell appear defective. Because $k$ and $s$ are correlated we
quote the fit-invariant local slope alongside $k$. We assessed adequacy from
the maximum absolute residual rather than a summary $R^2$; residual maxima are
of order \numrange{2e-4}{6e-4} units, so the fits are decompositions and not
smoothings. An independent verifier re-ran both campaigns using separate decks
and separate fitting code, and the two sets of coefficients agree to within
the reported residuals.

\subsection*{Provenance of the extracted netlists}

Geometric and connectivity checking is reported here only insofar as it
licenses reading the extracted netlists as an electrical description of the
layout the parameters come from.

\paragraph{Design-rule checking.} We ran two engines with deliberately
disjoint rule coverage and accepted a build only when both were clean:
KLayout with the shuttle's precheck deck in its three modes, and Magic with
the PDK deck inside the containerized flow the shuttle pins. The decks are not
interchangeable, since each carries rules the other omits, so a change neutral
under one can be considerably worse under the other.

\paragraph{Flattened, layer-merged geometry.} Hierarchical checking of the
same database produced counts that could not be interpreted, because Magic
closes the high-voltage implant layer per instance and reports slivers between
abutted cells whose implant is physically continuous. Flattening and merging
that layer before the check removed them, while KLayout reported none either
way, so the discrepancy is an artifact of hierarchical evaluation and not a
property of the layout. All results reported here are obtained on flattened
geometry with the relevant implant layers merged, and we quote hierarchical
violation counts nowhere.

\paragraph{LVS.} We ran LVS against reference netlists generated from the
design netlists, never from the layout, and derived the run mode per block
from the reference rather than fixing it, since a reference containing
subcircuit call cards requires hierarchical mode. We read verdicts from the
LVS database rather than from the one-line summary, because a circuit whose
subcircuit pins failed to pair is reported as \emph{skipped}, which is a
non-result indistinguishable from a pass in the log. The submitted database
matches on all 16 circuits with none skipped.

\paragraph{Parasitic extraction.} We extracted post-layout netlists in Magic
with both parasitic resistance and coupling capacitance, from the submitted
mask database, after flattening, applying two documented repairs: we merged
sub-ohm series resistors and dropped non-positive capacitors. For the single
cell this gives 61 devices, 4984 resistors and 1316 capacitors. For the
two-cell assembly we flattened the routed top level and extracted it \emph{in
one piece} rather than by stitching per-block extractions, giving 410 devices,
\num{10971} resistors and \num{3174} capacitors totaling
\SI{7.72}{\pico\farad}; only the digital configuration block was excluded, by
deleting its instance and driving its outputs as fixed levels. We verified
that the extracted netlists were the correct artifact by comparing the device
census against the LVS reference class by class, on card count and on summed
device width. We extracted no assembly deeper than two cells, and we present
no claim as post-layout for a deeper chain.

\note{The energy law: block budget and derivation}{si:s05}

Where a cell's charge goes, block by block, over the settling window they share. The main text quotes the ratio; this is the accounting behind it.

That second number is worth walking through rather than accepting, and
Table~2 of the main text splits it block by block over the one
\SI{402}{\nano\second} settling window they all share. The four servo
amplifiers take \SI{29.53}{\pico\joule} of it, \SI{74.5}{\percent}, and
they evaluate nothing: they hold the junctions' collectors where the
reference put them. The bias network takes \SI{8.70}{\pico\joule},
\SI{22.0}{\percent}, supplying the operating currents. The two output
mirrors take \SI{1.39}{\pico\joule}, \SI{3.5}{\percent}, to copy the
answer to the next cell. The four bipolar junctions that evaluate
$\mathrm{eml}$ take \SI{0}{\pico\joule} from the \SI{3.3}{\volt} rail,
because they are not on it: their \SI{3.962}{\micro\ampere} arrives
through the pedestal legs at \SI{1.2}{\volt}, and the
\SI{13.0}{\femto\joule} above is what it costs to move that current onto
the core's own \SI{47.9}{\femto\farad} of extracted capacitance. Three
of those four numbers are scaffolding and the fourth is the computation.
Divide the fourth into the total and the ratio falls out.

\paragraph{What shared scaffolding would be worth.} Table~2 of the main text
splits the cell's \SI{29.8686}{\micro\ampere} of supply current into a part every cell must replicate, the bias network at \SI{6.5605}{\micro\ampere} and the output mirrors at \SI{1.0494}{\micro\ampere} for \SI{7.61}{\micro\ampere} together, and the four servo amplifiers at
\SI{22.2587}{\micro\ampere}, which exist to hold operating points and are in
principle shareable across cells holding the same ones. Divided by the
\SI{1.98}{\micro\ampere} settling branch, half the \SI{3.96}{\micro\ampere}
servo tail, that gives $F(N) = \num{3.84} + \num{11.24}/N$ for scaffolding
amortized over $N$ cells. It reproduces the shipped
$F = \num{29.8686}/\num{1.98} = \num{15.085}$ at $N = 1$, which is the check
that the split is the right one, and falls to \num{5.25} at $N = 8$,
\num{4.19} at $N = 32$ and \num{3.84} in the limit. Taken with a per-cell
$C$ falling from \SI{2.00}{\pico\farad} toward parasitics, the product $FC$
improves by \num{115} at $N = 8$ and \num{144} at $N = 32$. Every figure
here is arithmetic on the energy law; no circuit was designed, and the
caveats in the main text apply.

\note{Bias-current dependence and the log-domain comparison}{si:s06}

Two questions a reader of the translinear literature reaches for: whether a lower bias current recovers the difference, and how a cell dissipating microwatts squares with subthreshold parts that dissipate nanowatts. Both are answered by the same cancellation, and neither changes a conclusion in the main text.

Lowering the bias current does not recover the difference. Below
\SI{250}{\nano\ampere} the predicted cancellation holds and energy is
bias-invariant; above it the settling time floors, because the
exponentiating path runs through a \SI{51.5}{\kilo\ohm} poly resistor
into \SI{1}{\pico\farad} and the resulting \SI{51.5}{\nano\second} time
constant, not the transistor, sets the floor. The shipped part sits above
the knee, so extra current buys no speed; moving below it reaches
\SI{19.3}{\pico\joule} per cell at \num{2.4} times the latency. That
figure is still \num{1.4} times an entire digital datapath, for one of
the seven cells the task needs. There is no bias current at which this
fabric is cheap.

That floor is not a sizing accident, and the reason is worth separating from
the number. $R_u$ is not a free parameter: the exponent scale is the single
product $\nu = R_u I_{\text{unit}} = \SI{26.9}{\milli\volt}$, which fixes one
unit of input current as one thermal-voltage e-fold, so $R_u$ follows from
the unit current and the specification rather than from a sizing choice. The
passive time constant is therefore $\tau_R = \nu C / I_{\text{unit}}$. It does move, since doubling the unit current halves it, but power moves with the unit current in the same proportion, so the product is
$\nu C = \SI{26.9}{\femto\joule}$ and is invariant to that trade. The
passive floor is thus the same cancellation the energy law describes,
reappearing in a resistor rather than in a transistor, and not an escape from
it. Two things would move it and neither is an amplifier: a smaller node
capacitance, which is parasitics, or a different exponent scale, which is a
specification. In particular no choice of servo topology moves this floor,
because the floor is not in the servo.

The same cancellation settles a comparison a reader of the log-domain
literature will reach for. Subthreshold circuits of comparable function run at
nanoamperes and dissipate
nanowatts~\cite{sarpeshkar2010ultralow,mandal2009logdomain}, orders of
magnitude below the per-cell power reported here, which invites the reading
that this cell is profligate. That comparison is about speed and not
about efficiency. Power falls with bias current and settling time rises in the
same proportion, so the quantity the composability overhead divides is energy
per evaluation, and the cancellation that holds it fixed is a property of the
energy law rather than of any sizing choice. Biasing a cell of this topology
into the nanoampere regime buys latency, not efficiency, and it pays the
overhead more slowly rather than paying less of it.

\note{The precision regime: Sarpeshkar's crossover and $kT/C$}{si:s07}

The arithmetic behind the statement that the fabric operates nine to eleven bits inside the regime where analog computation is supposed to win, and the noise-floor hypothesis that fails in the same direction.

So which of the two is wrong? Neither. The two statements are consistent
because they are about different objects, and it is the reading laid on
top of Sarpeshkar that fails, not his analysis. He priced a computing
element, and on that object we confirm him: the bare translinear core
beats the modeled datapath by \numrange{4}{134}$\times$, deep inside the
regime he identified. What the citation practice adds, and what he did
not claim, is that a machine assembled from such elements inherits the
advantage. It does not. The fabric pays for a composable, addressable,
individually calibrated element, and that surcharge is the
\num{3046}$\times$. The precision argument is therefore not wrong and not
sufficient. A low-precision regime is necessary for an analog fabric to
be competitive and is nowhere near enough, because the term that decides
the outcome is not in the precision--energy trade at all. The same
accounting explains why the fabric as built would need $b = \num{60.2}$
bits before its energy per evaluation undercut digital, a precision
\num{54} bits beyond its own mismatch ceiling.

The kT/C hypothesis fails in the same direction and confirms the
diagnosis. At the fabric's realized precision, thermal noise on a
sampling capacitor would permit capacitances in the attofarad range; the \SI{2}{\pico\farad} present is roughly $10^{6}$ times larger
than noise requires, and kT/C does not bind until \num{14.4} bits. What
sets $C$ is stability and parasitics, and what sets precision is
mismatch. Neither is a noise-floor phenomenon, and neither is reachable
by the arguments usually brought to analog efficiency.

\note{Alternative operators and the scope of the Hardy-field theorem}{si:s08}

What the exclusion theorem does and does not say once the domain is bounded, and the three operator variants tested behaviourally against $\mathrm{eml}$. No variant operator was designed as a circuit, so every cost here is an estimate from the shipped cell's current budget.

The scope of that theorem must be stated, because it is easy to claim more from it than it gives. It concerns germs at $+\infty$, while the
benchmark evaluates on bounded sampling boxes, and on a compact domain the
closure of $\{1,x\}$ under $\exp$, $\ln$ and the algebraic operations is
dense in the continuous functions. The theorem therefore does not say that a
trigonometric target cannot be approximated to a stated tolerance on a bounded
box. It says the grammar does not contain it. The practical claim is a
separate one and rests on experiment rather than on the theorem: the
$\sin(k \ln x)$ results above, where the fabric learns $k = 3$ and fails at
$k = 5$ and $k = 8$ at every depth and budget attempted. We keep the two
apart throughout. The \num{18} of \num{100} count belongs to the exact
statement, the fitted results belong to the empirical one, and neither is
offered as evidence for the other.

One claim in the opposite direction has been weakened by held-out testing and
should not be repeated in its strong form. Operators that place $\sin$ at a
port do reach $\sin$ trivially, because they contain it; what does not survive
is the broad coverage that made them look attractive. Scoring each target on a
grid disjoint from the one used to select the expression, $\sinh(u)-\sin(v)$
degrades from a worst-case error of \num{0.151} in sample to \num{22.6} out of
sample, with a median out-of-sample error of \num{0.574}, and the
diode--junction hybrid $u - W(e^{u+v}) - \sin(v)$ degrades on $\sin$ itself
from \num{3.9e-2} to \num{4.2e-1}. Apparent coverage close to the
enumeration's class cap is largely fitting. By this measure $\mathrm{eml}$,
which enumerates to far fewer classes, degrades least of the ten operators
tested (\num{0.673} to \num{1.56}), which is evidence that the coverage metric
rewards fitting rather than representation.

Three behavioral studies ask whether a larger or a different operator escapes
these limits. None involved a circuit-level netlist.

Adding a third, current-summed port, so the cell computes
$\exp(u)-\ln(v+s)+w$, is the natural way to buy addition and is nearly free in
the netlist. It bought nothing: at a matched \num{12} cells, \num{0} of
\num{6} targets moved from depth $\ge 3$ to depth $\le 2$, and at equal depth
the extended cell was worse by more than \SI{10}{\percent} in \num{30} of
\num{48} cases against better in \num{6}. The reason is structural. The fabric
already sums, since every port is affine in its inputs and its children's
outputs, so addition was never the missing capability. Depth here is limited
by mismatch, not by expressiveness, and buying expressiveness does not buy
depth.

Replacing $\mathrm{eml}$ with $\exp(u)-\operatorname{asinh}(v)$ or
$\sinh(u)-\operatorname{asinh}(v)$ removes two of its four conditioning
defects at no circuit cost: the out-of-domain fraction at depth $\le 4$ falls
from \num{0.625} to \num{0.000}, and port sensitivity goes from a pole at
$v \rightarrow 0^{+}$ to a bound of unity. They fix nothing else. Dynamic
range worsens by an order of magnitude, which is the wrong direction for a
fabric whose difficulty is fitting values inside a device span, and
trigonometry stays unreachable for the reason just given. Their completeness
is \emph{conjectured and not proved}, and the distinction matters because
completeness is the entire basis on which $\mathrm{eml}$ was proposed. What is
proved is that the ingredients transfer, since $\ln$ and $\exp$ can be
exhibited inside both grammars, and sufficient ingredients are not a
completeness argument. Odrzywo\l{}ek's theorem remains the only proved
completeness result in the family, so this is a trade and not an upgrade.

Looking outside translinear electronics, the strongest candidate is an
operating point rather than a transfer function: a diode in series with a
resistance settles to $u - W(e^{u+v})$ for the Lambert $W$. It is better
conditioned on every axis tested and is provably nonexpansive, since
$|\partial f/\partial u|+|\partial f/\partial v| = 1$ identically. Its
completeness is likewise not established, and it does not remove the
scaffolding, because a diode's operating point still has to be held. It makes
the scaffolding's job smaller, and whether a relaxed precision requirement
converts into a smaller $F$ is not answered here.

\note{CORDIC, lookup tables, and the per-task comparisons}{si:s09}

Two comparisons in which the fabric appears to do better than the main text reports, each stated with the half that removes the appearance.

Two apparent counter-results should be reported with both halves stated.
The fabric's narrowest defeat is against CORDIC at matched precision,
where it loses by \num{1.11}$\times$ at the published assumptions, inside the \SI{\pm 19}{\percent} parasitic spread, and wins in \num{27} of the \num{72} sensitivity corners. It loses that regime
anyway, because against the baseline a designer would build for a scalar exponential, a \num{16}-entry lookup table with interpolation,
the fabric loses by \num{8}$\times$ for a single cell and
\num{12}$\times$ for the seven-cell fabric. Separately, one energy
comparison shows the fabric using a tenth of the area of a lookup table
in five dimensions, and the lookup-table blow-up in high dimension is
real; but that comparison is a straw man, because digital resorts to a
table only when no closed form exists, and the closed forms in question
evaluate in tens of picojoules. The fabric wins nothing that a designer
choosing between real options would give it.

\note{The restoration proposition: Langevin landscapes and Kramers}{si:s10}

The numerical check on both halves of the proposition, and the price of a barrier that holds a value for a stated time. The proposition itself, and the fact that it is not ours, are in the main text.

Overdamped Langevin simulation of five canonical landscapes agrees on both
halves. The flat, Goldstone-type landscape gives
$\mathrm{d}\log\mathrm{Var}/\mathrm{d}\log t = +\num{0.997}$ against the
free-diffusion value of $+1$, so the held value diffuses and is lost; every
landscape that does restore destroys the value it was given, collapsing four
distinct held values into the landscape's own minimum. This corrects a claim
an earlier version of this argument made. A chemical steady state does restore, at
$+\num{0.007}$, but the value it holds is the one mass action puts there,
fixed by the rate constants and the feed. It holds \emph{its} value, not an
arbitrary one, and so falls to the proposition rather than exempting itself
from it.

The conclusion is stronger than the one it replaces. If restoration is
available only at isolated points, any substrate that holds a value holds one of finitely many, and a $b$-bit analog value needs $2^{b}$ wells,
which is to say the substrate has rebuilt digital restoration under another
name. A Kramers estimate prices them: holding one value for a time $t$ against
$kT$ at \SI{300}{\kelvin} needs a barrier of at least $kT\ln(t/\tau_0)$, which
for $\tau_0 = \SI{1}{\pico\second}$ is \num{6.9}\,$kT$ at \SI{1}{\nano\second}
and \num{27.6}\,$kT$ at one second. The composability overhead is therefore
not $\mathrm{eml}$'s fault, not an artifact of a \SI{130}{\nano\meter}
process, and not removable by swapping substrates. It is what is charged for
asking matter to hold a value that matter has no continuous mechanism for
holding.

\paragraph{Non-volatile analog memory is an instance, not an exception.}
A memristive memory holds a programmed conductance for weeks with no supply,
which is the strongest apparent counterexample to the proposition. It is not
one. Retention is bought with an energy barrier, so the states that survive
are discrete and few; programmed conductances relax and drift between them,
and the reliability literature puts open-loop analog storage at roughly four
bits~\cite{zhao2020reliability}. That is the proposition's own prediction. A substrate that holds a value holds one of finitely many, with $b$ small. The distinction worth keeping is between holding a
value \emph{in place}, which such a device does for free, and holding one
\emph{for the duration of an evaluation} against a driven input, which is
what this fabric pays for on every cell.

\note{Limits on the evidence}{si:s11}

The itemised inventory of what this study did and did not cover. The main text states every limit named here; this note is the detail behind each. The design is in fabrication but no part has returned, and no number in either document measures a physical device.

Post-layout extraction covered a single cell and a two-cell assembly and
nothing deeper. Every three- and four-cell result is schematic-level with
respect to interconnect, and no extracted evidence exists at the chain
depths the fabric analysis uses; the \SI{19}{\percent} parasitic
inflation obtained on the two-cell assembly is applied to deeper
configurations as a modeling assumption. Within the two-cell assembly, one corner, \texttt{ss} at \SI{-40}{\celsius}, failed to integrate
before the stimulus fired and is reported as unavailable rather than as a
pass. The extracted supply mesh carries no series resistance, so supply
IR drop is absent from every post-layout result.

The $u$-port coupling is absent from the design entirely. The netlist
generator wires every coupling weight to the next cell's $v$ port and
holds every downstream $u$ port at a fixed pedestal; the coupling exists
as a label with no instance behind it, and the simulated $u$ swing there
is \num{0.000000} units in every weight configuration at every cell count
built. Every $u$-port quantity in this paper is composed from the single-cell fit and the simulated
device span rather than swept end to end. Composition results are
likewise positive-coupling only: at negative coupling the logarithm's
argument reaches $-\num{22.5}$ units, where it is not evaluable, and only
2 of 16 negative weight codes survive at either link simulated.

The loop-gain analysis was never completed. Loop gain and phase margin
were obtained only on the schematic-sizing netlist, and no loop-gain
probe was run against any extracted netlist. The post-layout stability
evidence is decay under an injected kick, which is a weaker statement
than a simulated margin, and the \num{2}\,pF compensation capacitor whose
cost dominates the energy law is justified by that weaker evidence
together with the observation that an uncompensated variant rings in
\num{15} of \num{25} corners.

The depth-3 and depth-4 configurations that carry the behavioral results
sit below the shipped silicon's own mismatch floor. Per-cell resolution
on the extracted part is \num{2.28} bits, and no uncalibrated
configuration reaches the \SI{1}{\percent} tolerance column at any depth
in the corpus study; that column is exactly \num{0.000} in every
uncalibrated row. Accuracy projections at those depths therefore describe
a machine that, built on this silicon, would produce no signal. They
bound what the architecture could do given a cell that does not exist,
and they should be read that way.

All fabric energy figures are lower bounds. Weight-DAC energy was
simulated at zero code only, so the switching and static cost of the
coupling and input DACs at the non-zero codes any trained configuration
would use is absent from every fabric energy number reported. The
composability overhead, the fabric-versus-digital ranges and the regime
analysis all move in the same direction, against the fabric, if that energy is added.

Several inputs to the comparison are models rather than measurements and
are flagged as such throughout. The digital baseline is an
energy-per-operation model~\cite{horowitz2014computing} scaled from
\SI{45}{\nano\meter} to \SI{130}{\nano\meter} and multiplied by an
assumed control overhead; the sensitivity box exists because those two factors are assumptions, and the ranges rather than the point
estimates are what should be cited. The subthreshold slope factor $n$ is
a typical value, not an extraction, and it enters the energy law
linearly. The per-stage latency that sets the seven-cell and
\num{32}-cell energies rests on two simulated points and is flagged weak
in the underlying record. Two per-cell power figures appear in the source
data and each is correct in its own context, so every use above names
which it takes: \SI{98.566}{\micro\watt} is one isolated cell in die
context and is the basis of the \SI{39.6}{\pico\joule} per cell and
hence of the composability overhead, while \SI{96.95}{\micro\watt} is the
fitted per-cell slope of the two-cell in-chain power law and is the basis
of the seven-cell and \num{32}-cell energies. They differ by
\SI{1.7}{\percent} because the in-chain slope absorbs shared bias
distribution that an isolated cell pays for alone. The AI Feynman equation set used here was
transcribed offline and its variable sampling ranges are documented
substitutes for the published ones, verified for structure, dimensional
homogeneity and sampler finiteness, with independent physics checks on
\num{54} of the \num{100} equations; the remaining \num{46} rest on
structure and dimensions alone. The trigonometric exclusion count is
robust to this substitution, since it depends on the equations' operator
sets and not on their ranges, but the baseline accuracy figures are not.

Finally, the network-scale results are modeled rather than simulated at
circuit level. Corpus fractions and the trim exchange rate come from a
behavioral cell evaluated in double precision, no network larger than
four cells was simulated at circuit level, and nothing here tests whether
an eighteen-cell network settles or is stable. The area scaling law rests
on a single placed point at $N = 2$; nothing larger than two cells was
placed and routed, and the rescaling stages the depth analysis calls for
are a circuit that was never designed.

Three further limits belong to the extension studies. The restoration proposition above is proved for autonomous
dynamics and therefore covers storage; it does not cover driven composition,
which is what this fabric does, and we have not shown that the impossibility
carries across. The two better-conditioned operators are demonstrated to fix
$\mathrm{eml}$'s domain hole and its unbounded $v$-port sensitivity, and their completeness is conjectured only. The identities that transfer $\exp$ and $\ln$ into their grammars are proved, but sufficient ingredients are not
completeness, and the same holds for the Lambert-$W$ operating point. The
third-port and operator-comparison results are behavioral: no variant
operator was designed as a circuit, laid out, or simulated at circuit level,
so their power figures are cost estimates from the cell's own current budget
rather than extracted quantities.

\note{The random-feature regime: protocol and per-system results}{si:s12}

The protocol behind Section~2.7 of the main text, and the tables the medians
there are drawn from.

\subsection*{What was run}

The interior weights of a fabric are drawn once and frozen. Every cell output
is treated as a feature, and only a linear readout is fitted, by ridge
regression whose penalty is chosen by generalized cross-validation over a
\num{73}-point logarithmic grid. Features are standardized on the training set
and railed at eight standard deviations before the readout sees them, which
models a finite output-stage input range; the fraction of entries railed is recorded per run and is \num{0.000} in every row reported here, so
the rail is immaterial to these results.

This is an extreme learning machine, not a reservoir. The fabric is
feedforward and has no fading memory, and the recurrence in the closed-loop
column belongs to the differential equation rather than to the device. We use
the term ELM throughout and make no claim about tasks with memory.

Everything else is the matched study's, imported rather than reimplemented:
the same \num{16} systems, initial conditions, training box, input offset and
\num{768} uniform samples; the same closed-loop substitution, blow-up guard
and \textsc{nrmse} definitions; best-of-seeds selected on \emph{training}
error; and \texttt{noisy=False} at evaluation, so read noise is absent from
this study exactly as it is absent from the trained one. Only the fitting rule
differs.

Two protocol differences favor the frozen fabric and should be stated whenever
these numbers are quoted. It ran five seeds against the trained study's three.
And for the one two-output system, \texttt{enzyme\_cascade}, it shares a single
feature bank across both outputs where the trained study fits a separate fabric
per output. Neither approaches the effect sizes reported.

Three hardware configurations are used: \emph{ideal}, real-domain arithmetic;
\emph{pedestal}, the realized cell's $\ln(v+s)$ with $s = \num{2.547}$; and
\emph{silicon}, the pedestal together with the extracted mismatch, a
\num{30}-unit output rail and \num{8}-bit weights. The per-hop $v$-port
attenuation is off in all three, being derived from the pedestal, so enabling
both would double-count the same physics.

\subsection*{The control that settles what freezing is worth}

A frozen fabric reading all its cells differs from the trained fabric in two
ways at once: the interior is not trained, and the readout is wider. Holding the second fixed, by freezing the interior but reading layer~0 only, which is the readout the trained fabric uses, leaves the two level: median ratios of
\numrange{0.82}{1.95} across the four matched pairs, the frozen variant winning
between \num{4} and \num{10} of \num{16}. Gradient descent through the
behavioral model is therefore not what limits the trained fabric, and the
median \num{21.5}$\times$ reported in the main text is not an optimization
artifact. The order-of-magnitude improvement comes from the wider readout, and
an all-cell readout costs an output path per cell.

\subsection*{Per-system pointwise fit on the validation box}

\begin{center}\small
\begin{tabular}{lrrrrr}
\toprule
system & trained & frozen & matched & polynomial & $\tanh$ \\
 & silicon & silicon & MLP & & ELM \\
\midrule
\texttt{hill\_n1} & 0.0262 & 0.0098 & 0.0032 & 0.0158 & 0.0013 \\
\texttt{hill\_n2} & 0.0250 & 0.0017 & 0.0006 & 0.0038 & 6.6e-05 \\
\texttt{hill\_n4} & 0.0473 & 0.0044 & 0.0004 & 0.0100 & 4.6e-07 \\
\texttt{enzyme\_cascade} & 0.0726 & 0.0259 & 0.0069 & 0.0197 & 0.0092 \\
\texttt{gene\_autoreg} & 0.0271 & 0.0007 & 0.0010 & 0.0016 & 6.2e-05 \\
\texttt{monod\_chemostat} & 0.0534 & 0.0208 & 0.0097 & 0.0158 & 0.0102 \\
\texttt{gompertz} & 0.0274 & 0.0097 & 0.0060 & 0.0139 & 0.0021 \\
\texttt{bimolecular} & 0.0758 & 3.6e-08 & 0.0060 & 3.6e-16 & 8.1e-08 \\
\texttt{gene\_repressilator} & 0.0610 & 0.0301 & 0.0018 & 0.0904 & 0.0001 \\
\texttt{goodwin} & 0.0733 & 0.1662 & 0.0034 & 0.2805 & 0.0055 \\
\texttt{lotka\_volterra} & 0.0833 & 2.0e-08 & 0.0058 & 2.9e-16 & 0.0020 \\
\texttt{logistic} & 0.0207 & 5.4e-09 & 0.0051 & 1.0e-15 & 8.1e-09 \\
\texttt{duffing\_damped} & 0.2196 & 1.1e-07 & 0.0023 & 3.4e-16 & 5.2e-08 \\
\texttt{pendulum\_damped} & 0.2466 & 5.8e-06 & 0.0017 & 3.4e-06 & 1.7e-08 \\
\texttt{brusselator} & 0.1631 & 1.3e-07 & 0.0121 & 1.8e-16 & 0.0065 \\
\texttt{mm\_bisub} & 0.1520 & 0.1663 & 0.0455 & 0.1387 & 0.0438 \\
\midrule
median & 0.0668 & 0.0030 & 0.0042 & 0.0069 & 0.0007 \\
\bottomrule
\end{tabular}\end{center}

\subsection*{Per-system closed-loop trajectory}

\begin{center}\small
\begin{tabular}{lrrrrr}
\toprule
system & trained & frozen & matched & polynomial & $\tanh$ \\
 & silicon & silicon & MLP & & ELM \\
\midrule
\texttt{hill\_n1} & 0.0370 & 0.0115 & 0.0040 & 0.0174 & 0.0015 \\
\texttt{hill\_n2} & 0.0373 & 0.0021 & 0.0003 & 0.0048 & 4.6e-05 \\
\texttt{hill\_n4} & 0.0554 & 0.0046 & 0.0007 & 0.0091 & 4.6e-07 \\
\texttt{enzyme\_cascade} & 0.1194 & 0.0461 & 0.0179 & 0.0442 & 0.0190 \\
\texttt{gene\_autoreg} & 0.0062 & 0.0002 & 2.9e-05 & 0.0009 & 1.2e-06 \\
\texttt{monod\_chemostat} & 0.0447 & 0.0260 & 0.0022 & 0.0109 & 0.0032 \\
\texttt{gompertz} & 0.0056 & 0.0011 & 0.0007 & 0.0016 & 3.1e-05 \\
\texttt{bimolecular} & 0.0918 & 2.0e-07 & 0.0058 & 1.8e-07 & 1.7e-07 \\
\texttt{gene\_repressilator} & 0.1231 & 0.0537 & 0.0022 & 0.3038 & 5.3e-05 \\
\texttt{goodwin} & 0.1356 & 0.2283 & 0.0127 & 0.6015 & 0.0135 \\
\texttt{lotka\_volterra} & 0.5751 & 3.9e-06 & 0.1284 & 4.2e-06 & 1.8e-05 \\
\texttt{logistic} & 0.0100 & 1.0e-07 & 0.0019 & 1.0e-07 & 1.0e-07 \\
\texttt{duffing\_damped} & 1.0698 & 1.6e-06 & 0.0169 & 1.3e-06 & 1.4e-06 \\
\texttt{pendulum\_damped} & 0.2150 & 8.9e-06 & 0.0019 & 4.1e-06 & 4.4e-07 \\
\texttt{brusselator} & 0.7214 & 2.1e-06 & 0.1198 & 1.2e-06 & 5.8e-06 \\
\texttt{mm\_bisub} & 0.0572 & 0.0208 & 0.0035 & 0.0129 & 0.0016 \\
\midrule
median & 0.0745 & 0.0016 & 0.0029 & 0.0032 & 2.5e-05 \\
\bottomrule
\end{tabular}\end{center}

\end{document}